\documentclass[aps,prb,preprint,groupedaddress,showpacs,showkeys,amsmath]{revtex4-1}
\usepackage[dvips]{graphicx}
\usepackage{dcolumn}
\usepackage{bm}
\usepackage{color}
\usepackage{longtable}
\usepackage{ctable}
\usepackage{rotating}
\begin{document}

\title{The role of van der Waals  and exchange interactions in high-pressure solid hydrogen}

\author{Sam Azadi}

\affiliation{Royal School of Mines and the Thomas Young Centre, Imperial College London, SW7 2AZ London, 
School of Physics and Centre for Science at Extreme Conditions,
University of Edinburgh, Edinburgh EH9 3JZ, United Kingdom}

\email{s.zadi@ic.ac.uk ; sam.azadi@ed.ac.uk}

\author{Graeme J. Ackland}

\affiliation{School of Physics and Centre for Science at Extreme Conditions,
University of Edinburgh, Edinburgh EH9 3JZ, United Kingdom}

\date{\today}

\begin{abstract}
We investigate the van der Waals interactions in solid molecular
hydrogen structures.  We calculate enthalpy and the Gibbs free energy
to obtain zero and finite temperature phase diagrams, respectively.
We employ density functional theory (DFT) to calculate the electronic
structure and Density functional perturbation theory (DFPT) with van
der Waals (vdW) functionals to obtain phonon spectra.  We focus on the
solid molecular $C2/c$, $Cmca$-12, $P6_3/m$, $Cmca$, and $Pbcn$
structures within the pressure range of 200 $<$ P $<$ 450 GPa. We propose
two structures of the $C2/c$ and $Pbcn$ for phase III which are stabilized
within different pressure range above 200 GPa.
We find that vdW functionals have a big effect on
vibrations and finite-temperature phase stability, however, different
vdW functionals have different effects.  We conclude that, 
in addition to the vdW interaction, a correct treatment of
the high charge gradient limit is essential.  
We show that the dependence of molecular bond-lengths on
exchange-correlation also has a considerable influence on the calculated
metallization pressure, introducing errors of up to 100GPa.
\end{abstract}

\maketitle

\section{Introduction}

Determining the phase diagram of high-pressure hydrogen is one of the great
challenges of condensed matter physics.  Since 1935, when it was predicted that
molecular solid hydrogen would become a metallic atomic crystal at 25
GPa\cite{1935} high-pressure hydrogen has been studied intensively 
by theory and experiment. It was also predicted theoretically the possible
existence of room-temperature superconductivity\cite{Ashcroft} and metallic
liquid ground state\cite{Bonev}. Additional interests rise from 
the relevance of solid hydrogen to astrophysics\cite{Hemley,Ginzburg}. 

Early infrared (IR) and Raman measurements at low temperature
suggested the existence of three solid-hydrogen
phases\cite{Hemley}. Phase I, which is stable up to 110$\pm$5 GPa, is
a molecular solid composed of quantum rotors arranged in a hexagonal
close-packed structure.  Phase I spans a wide pressure-temperature (P-T)
range. Hence the physical properties of phase I of hydrogen evolve extensively 
as the solid becomes nine times denser. It has been accepted that the melting 
curve of hydrogen exhibits a maximum below 130$\pm$10 GPa at around 
1000$\pm$100 K\cite{Gregoryanz03,Kechin,Eremets09,Howie3}. 
Extrapolating the existing data to pressures larger than 250 GPa
predicts room temperature melting at P $>$ 300$\pm$50 GPa, but
thermodynamics requires that the melt line will become shallower above
the high-entropy phase IV\cite{Magdprb17}.  Changes in the
low-frequency regions of the Raman and infrared spectra imply the
existence of phase II, also known as the broken-symmetry phase, above
110$\pm$5 GPa.  Phase II is observed at temperatures below 100$\pm$20K.
 The appearance of phase III at 150 GPa
 and below room temperature is accompanied by a large
discontinuity in the Raman spectrum and a strong rise in the IR spectral
weight of molecular vibrons\cite{Kohanoff99,loubey}.  Phase IV, characterized
by the two vibrons in its Raman spectrum, was recently discovered at
300 K and pressures above 230 GPa\cite{Eremets, Howie,
  Howie2}. Another new phase has been found at pressures above 200 GPa
and higher temperatures (for example, 480 K at 255
GPa)\cite{Howie3}. This phase is thought to meet phases I and IV at a
triple point, near which hydrogen retains its molecular character. The
most recent experimental results\cite{Simpson} indicate that H$_2$ and
hydrogen deuteride at 300 K and pressures greater than 325 GPa
transform to a new phase V, characterized by substantial weakening of
the vibrational Raman activity. Other features include a change in the
pressure dependence of the fundamental vibrational frequency and the
partial loss of the low-frequency excitations.

Although it is very difficult to reach the hydrostatic pressure of more than
400 GPa at which hydrogen is normally expected to metallize, some experimental
results have been interpreted as indicating metallization at room temperature
below 300 GPa\cite{Eremets}. However, other experiments show no evidence of
the optical conductivity expected of a metal at any temperature up to the
highest pressures explored\cite{zha}. Experimentally, it remains unclear
whether or not the molecular phases III and IV are metallic, although it has
been suggested that phase V may be non-molecular (atomic)\cite{Simpson}.
Metallization is believed to occur either via the dissociation of hydrogen
molecules and a structural transformation to an atomic metallic
phase\cite{samprl,Eremets}, or via band-gap closure within the molecular
phases\cite{MStadele,KAJohnson}. In this work we investigate 
the influence of van der Waals interactions on the metallization 
by performing finite temperature phase diagram calculations for 
insulator and metallic molecular structures. 

The phase diagram of high-pressure solid hydrogen has mainly been
investigated using density functional theory (DFT) with local and
semi-local exchange-correlation (XC) functionals
\cite{cogent,Pickard,Pickard2,Goncharov,Magdau,Naumov,Morales2013,Clay,JETP,singh,PRB13}.
In particular, DFT with generalized gradient approximation (GGA)
functionals has been widely applied to search for candidate low-energy
crystal structures and to calculate their vibrational properties.
Recently, DFT-GGA was used to reinterpret the IR spectrum of
hydrogen-deuterium mixtures in molecular structures\cite{dias2016new,Magdprl17},
and it has been found that the isotope effect leads to a completely
different spectroscopic signal in hydrogen-deuterium mixtures.  More
accurate quantum Monte Carlo methods\cite{Matthew1,samjcp15,sambenz}
are employed to calculate the static phase diagram\cite{NJP,Neil15}
and excitonic and quasi-particle band gaps for molecular
phases\cite{PRB17}. Just recently an interesting classical
thermodynamic model that reproduces the main features of the solid
hydrogen phase diagram has been introduced\cite{Magdprb17}.  It was
shown that the general structure types, which are found by electronic
structure calculations and the quantum nature of the protons, can also
be understood from a classical viewpoint.

The relative contribution of the van der Waals (vdW) interactions to
the cohesive properties of the various solid molecular structures of
high-pressure hydrogen has not been understood. First principles study
of ice phase diagram provides an important consequence, likely to be
of relevance to hydrogen-rich molecular crystals in general, which is
that transition pressures obtained from DFT-XC which neglect vdW
forces are greatly overestimated\cite{santra11}.  We have recently
studied the phase diagram of compressed crystalline benzene using
modern vdW and GGA density functionals\cite{benz16}.
 We found that the vdW forces play crucial role in 
prediction of phase stability and transition pressure in crystalline benzene. 
Considering the aforementioned results for ice and crystalline benzene, 
it may be expected that the vdW interactions are important in  
phase diagram calculations of low-Z hydrogen-dominant molecular 
crystals.
Similar to other rare
gases, simple H$_2$ molecules are weakly bounded due to vdW forces in
ambient conditions. A detailed study of the helium-nitrogen system in
a diamond-anvil cell using synchrotron X-ray diffraction, Raman
scattering and optical microscopy, indicates a novel class of vdW
compounds that are formed only at high
pressures\cite{vos92}. Theoretical study of liquid-liquid
insulator-metal-transition phase boundaries for high-pressure
deuterium\cite{knudson} predicts that the pressure-temperature phase
diagram results which are simulated by vdW functionals are in better
agreement with experiment comparing with conventional density
functionals.  Therefore, we believe that it is important to understand
the contribution of vdW interactions in static and dynamic phase
diagrams of high-pressure solid molecular hydrogen.

The main purpose of current work is to study the role of vdW forces in
the properties of molecular phases of high-pressure solid hydrogen.
We calculate enthalpy-pressure static phase diagram and also the Gibbs
free energy dynamic phase diagram up to room temperature.  We employ
two widely used vdW functionals of vdW-DF1\cite{vdw10,vdw11} and
vdW-DF2\cite{vdw2} and compare them with the results from conventional
DFT functionals.  Although these vdW functionals are tested on a broad
range of materials including traditional metals, ionic compounds, and
insulators \cite{Klimes11,Klimes12}, they were not employed before to
calculate the finite-temperatures phase diagram of high-pressure solid
hydrogen.  We consider five specific molecular structures with space
groups $P6_3/m$, $C2/c$, $Pbcn$, $Cmca$-12, and $Cmca$ within pressure
range between 200 to 400 GPa.  These structures were
  predicted by the {\it ab initio} random structure searching
  method\cite{Pickard}.  According to those calculations, the $C2/c$
  and $Pbcn$ structures are candidates for phases III and IV,
  respectively. The $C2/c$ structure includes weakly-bonded nearly
  graphene-like layers, while the $Pbcn$ phase adopts two different
  layers of nearly graphene-like three-molecule rings with elongated
  $H_2$ molecules and unbound $H_2$ molecules\cite{Pickard,Howie,
    Magdau}. The $Cmca$-12 structure is similar to $C2/c$ but slightly
  denser and has a much smaller metalization pressure\cite{NJP}. The
  $Cmca$ phase shows weaker molecular bonds than $C2/c$ and $Cmca$-12
  and is the only metallic molecular phase within the studied pressure
  range.  The structure of the $P6_3/m$ differs from the other layered
  phases. In this phase three quarters of the $H_2$ molecules lie flat
  in the plane and one quarter lie perpendicular to the plane.  More
  recently, several other structures involving small symmetry-breaking
  distortions from $Pbcn$ have been proposed for Phase IV, but
  molecular dynamics simulations at the temperatures where Phase IV is
  observed show that the unbound molecules rotate, increasing the
  time-averaged symmetry to
  $P6/mmm$\cite{liu2012room,liu2013proton,Magdau,Magdprb17}.  With the
  PBE functional, further small distortions mean that the lowest known
  energy candidate for phase II is $P2_1/c$
  \cite{pickard2009structures} and for phase III
  $P6_122$\cite{monserrat16} and $C2/c-24$
  \cite{pickard2009structures} below and above 200GPa
  respectively. With PBE the $Cmca$ phase is stabilised well into the
  pressure range where it has been ruled out experimentally.

Comparison between experiment and theory is typically done by
comparing at the same pressure.  However it is important to note that
neither experiment nor theory is very reliable in measuring pressures.
Experimental pressures are estimated with respect to the diamond
absorption edge, a scale which has been frequently revised by tens of
GPa\cite{akahama2010calibration}.  The natural variable for quantum calculations is
volume, with pressure being a calculated quantity.
Standard DFT codes calculate the differential of
the energy with respect to an affine rescaling of the simulation cell,
neglecting zero point contribution and the different compressibility
of inter- and intramolecular regions.  Consequently experimental and
theoretical measures are highly self-consistent, but comparing the two
is dangerous.

Crystal structure and hydrogen positions in the primitive unit-cell
are the fundamental inputs for {\it ab initio} phase diagram
calculations.  Due to lack of any established experimental structure
determination, there is no option but to use structures predicted by
DFT.  Most of the structures have been predicted by
Perdew-Burke-Ernzerhof (PBE)\cite{PBE} exchange correlation (XC)
functionals\cite{Pickard,Pickard2}  which has become the {\it de
  facto} standard in structure searching packages. It is now generally
accepted that DFT results for high-pressure hydrogen strongly depend
on the choice of exchange-correlation functional\cite{cogent,Clay,PRB13}
and although PBE performs well in identifying candidate structures, it
is poor at describing the relative energies of configurations and
the properties of the molecular
bond\cite{Clay}. The sensitivity to choice of DFT-XC functional
depends on the property being studied, and serious doubts about the
accuracy of the results persist. Is the pressure calculated correctly?
How do the interatomic interaction energy, the bond-stretch energy,
the phase diagram, the metallization mechanism, and the phonon
spectrum depend on the approximation used for the XC functional? How
accurate should we expect DFT calculations of measured quantities such
as infrared (IR) and Raman spectra to be? Answering these questions is
necessary to assess the reliability of the many existing DFT
simulations of high-pressure solid hydrogen. In this work, we examine
the accuracy of non-local vdW functionals in prediction of the
properties of high-pressure solid hydrogen.

The paper is organized as follows. Section \ref{CD} describes the details of
our vdW-DF calculations.  The static and dynamic phase diagrams and also 
calculated IR intensities are discussed in Sec.\ \ref{REDI}.
Section \ref{CON} concludes.

\section{Computational details}\label{CD}

Given that the energy differences between solid hydrogen molecular  
structures are small, the calculations must be performed with the
highest possible numerical precision.  Our DFT calculations were carried
out within the pseudopotential and plane-wave approach using the 
latest version of Quantum ESPRESSO suite of programs\cite{QS}. 
Our DFT calculations used non-relativistic norm conserving pseudopotentials
which were obtained by the Perdew-Burke-Ernzerhof (PBE)\cite{PBE}
exchange correlation functionals.
We used a basis set of plane waves with an energy cutoff 80 Ry. 
Geometry and cell optimisations employed a dense
$16\times16\times16$ ${\bf k}$-point mesh. The quasi-Newton
algorithm was used for cell and geometry optimisation, with convergence
thresholds on the total energy and forces of 0.01 mRy and 0.1 mRy/Bohr,
respectively, to guarantee convergence of the total energy to less
than 1 meV/proton and the pressure to less than 0.1 GPa/proton.

Vibrational frequencies and phonon spectra were calculated using density-functional
perturbation theory as implemented in Quantum ESPRESSO\cite{QS}. 
We use quasi-harmonic approximation to calculate the vibrational free energy\cite{Baroni}:

\begin{equation}
  F_{ph} (T,V) = k_{B} T \sum_{i, \bf{q}} ln \{ 2 sinh[\hbar \omega_{i, \bf{q}} (V)/2k_{B}T] \},
\label{eq2}
\end{equation}

where $k_B$, $V$, and $\omega_{i, \bf{q}}$ are Boltzmann constant, unit cell 
volume, and eigenvalue of the phonon Hamiltonian, respectively.
The zero point (ZP) pressure is included in our phase diagram calculations 
by $P_{ZP}=-(\partial E_{ZP}/ \partial V)$, where the $E_{ZP}$
per proton at a specific cell volume $V$ was estimated within the
quasi-harmonic approximation: $E_{\text{ZP}}(V) = \hbar
\overline{\omega}/2$, where $\overline{\omega} = \sum_{\bf q}
\sum_{i=1}^{N_{\text{mode}}} \omega_{i}({\bf q})/(N_{\bf q}
N_{\text{mode}})$. $N_{\text{mode}}$ and $N_{\bf q}$ are the
numbers of vibrational modes in the simulation cell and phonon wave vectors
${\bf q}$, respectively, and the summation over ${\bf q}$ includes all
${\bf k}$-points on a $2 \times 2 \times 2$ grid in the Brillouin zone.

Our electronic structure and lattice dynamic results are calculated
by vdW-DF1\cite{vdw10} and vdW-DF2\cite{vdw2} functionals. 
The Slater exchange and Perdew-Wang (PW)\cite{pw} 
correlation functionals are used in both vdW-DF1 and vdW-DF2
which means the correlation energy is approximated by 
local density approximation (LDA). In vdW-DF1 the gradient 
correction on exchange energy uses the revised version of PBE\cite{rpb}, 
whereas vdW-DF2 uses an optimized version of PW86\cite{pw86} 
which is named PW86R\cite{pw86r}. These functionals 
use different kernel for non-local energy term which
accounts approximately for the non-local electron correlation
effects. The non-local term is obtained using a
double space integration, which represents an improvement
compared to local or semi-local functionals, especially in the 
case of layered structures\cite{Rydberg03}.

\section{Results and discussion}\label{REDI}
\subsection{Static enthalpy-pressure phase diagram}
\begin{figure}
\begin{tabular}{c c}
\includegraphics[width=0.5\textwidth]{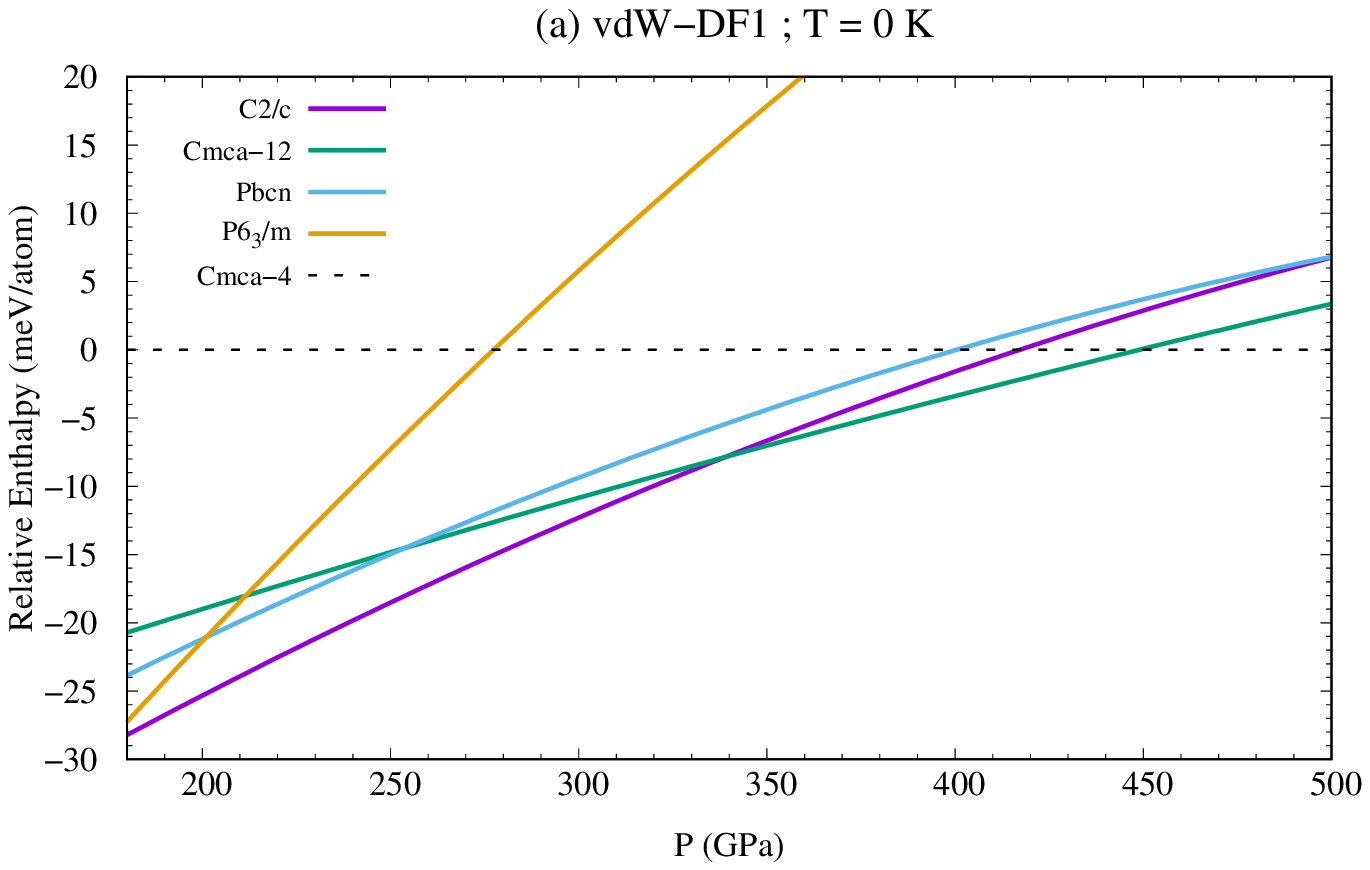}&
\includegraphics[width=0.5\textwidth]{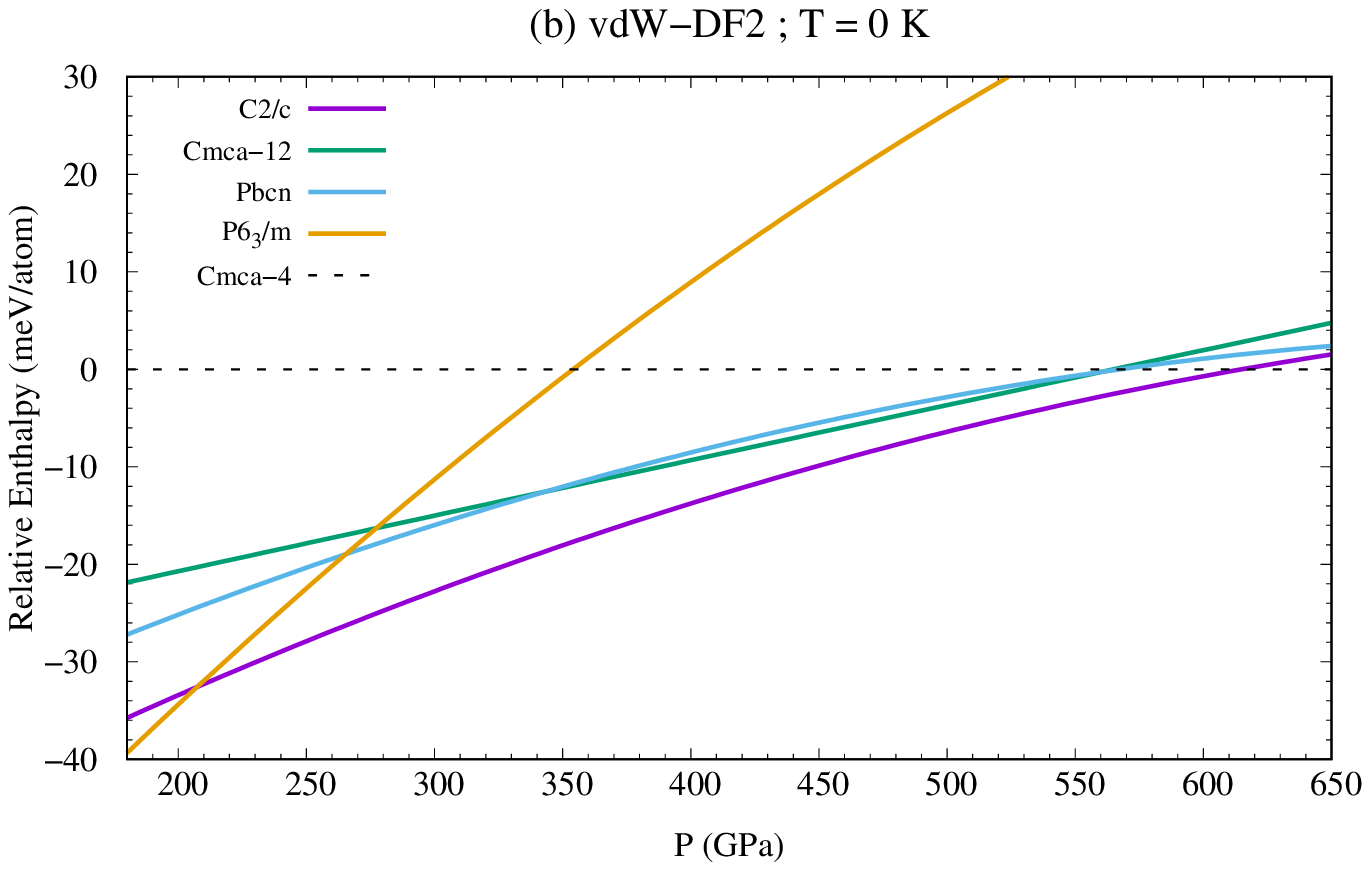}\\
\end{tabular}
\caption{\label{HP} (Color online) Relative enthalpy per atom as a function of
pressure calculated using two different vdW functionals of vdW-DF1 and vdw-DF2.
The static lattice (no phonon contributions) enthalpies of molecular crystal 
structures are presented relative to the enthalpy of the metallic $Cmca$ structure.} 
\end{figure}

Figure \ref{HP} illustrates static lattice enthalpy-pressure phase diagram 
calculated using the vdW-DF1 and vdW-DF2 functionals. According to our vdW-DF1 
results the $C2/c$, $Cmca-12$, and metallic $Cmca$ phases are stable in the 
pressure ranges $<$ 200-340, 340-450, and $>$ 450 GPa, respectively. Our vdW-DF2 
calculations predict that the $P6_3/m$ is stable below 210 GPa and the $C2/c$
is the most stable insulator phase until it transits to metallic $Cmca$ at pressure 
of 625 GPa. The relative stability of phases predicted by vdW-DF1 and vdW-DF2 
is not similar. The vdW-DF1 and vdW-DF2 functionals predict that the molecular
insulator to molecular metallic phase transition occurs at pressures 450, and 625 
GP, respectively. 
Our static enthalpy-pressure phase diagram obtained by vdW functionals 
differ from the previous results which are calculated by conventional 
DFT functionals\cite{PRB13}. PBE static lattice phase diagram predicts 
that the $P6_3/m$, $C2/c$,$Cmca$-12, and metallic $Cmca$ phases
are stable in the pressure ranges $<$ 110, 110–245, 245–370,
and $>$ 370 GPa, respectively. The semi-local Becke-Lee-Yang-Parr (BLYP)
functional\cite{BLYP} enthalpy-pressure phase diagram indicates that 
the $P6_3/m$, $C2/c$,$Cmca$-12, and metallic $Cmca$ phases
are stable in the pressure ranges of $<$ 160, 160–370, 370–430, and
$>$ 430 GPa, respectively. Calculations using Local density approximation and 
PBEsol\cite{PBEsol}  also give diverse properties\cite{cogent}.

It has been proven that DFT electronic structure results in the case
of high-pressure solid molecular hydrogen dramatically depend on the
XC functional. In our previous work\cite{PRB13}, we argued that the
self-interaction (XC-SI) error present in the XC functionals plays
crucial role in the study of $H_2$ systems. For instance, the XC-SI
errors of the LDA, GGA, and BLYP total energies of a single $H_2$
molecule are 1.264, -0.126, and 0.0846 eV,
respectively\cite{Polo}. These values are more than two orders of
magnitude larger than the conventional DFT enthalpy differences
between the crystal structures of high-pressure solid hydrogen.
Consequently DFT is highly dependent on cancellation of XC-SI errors:
this is reasonable when comparing different molecular structures

The XC energy in vdW functionals, in general, can be expressed as: 
$E_{XC} = E_{X}^{GGA}+E_{C}^{LDA}+E_{C}^{non-local}$
where the $E_{X}^{GGA}$ is revPBE\cite{rpb} and PW86R\cite{pw86r} 
for vdW-DF1 and vdW-DF2, respectively. The non-local part of 
correlation energy $E_{C}^{non-local}$ by definition does not 
suffer from the Coulomb self-energy of each electron. The local 
correlation energy $E_{C}^{LDA}$ is identical in vdW functionals considered 
in this work. Hence, the XC-SI errors of vdW-DF1 and vdW-DF2 are
mostly related to the X-SI errors. It was reported that\cite{Emmon} 
the PW86 functional, shows the  most  consistent  agreement  with  
exact-exchange Hartree-Fock (HF) interaction energies for $H_2$ clusters. 
In the next section we discuss the overall behaviour of vdW-DF1 and vdW-DF2 
for large density limit. It has been comprehensively discussed that 
properties such as molecular bond-length and interactions energies calculated 
by vdW-DF2 are improved\cite{Emmon}. However, at this stage we can not comment 
that this improvement is due to a lower SIE rather than other limits being obeyed. 
The study of SIE in DFT-XC functional demands a separate work.

To obtain a deep understanding of the static phase diagram and also
for comparing our results with experiment, we calculate
pressure-density equation of state using vdW-DF1 and vdW-DF2.
\begin{figure}
\begin{tabular}{c c}
\includegraphics[width=0.45\textwidth]{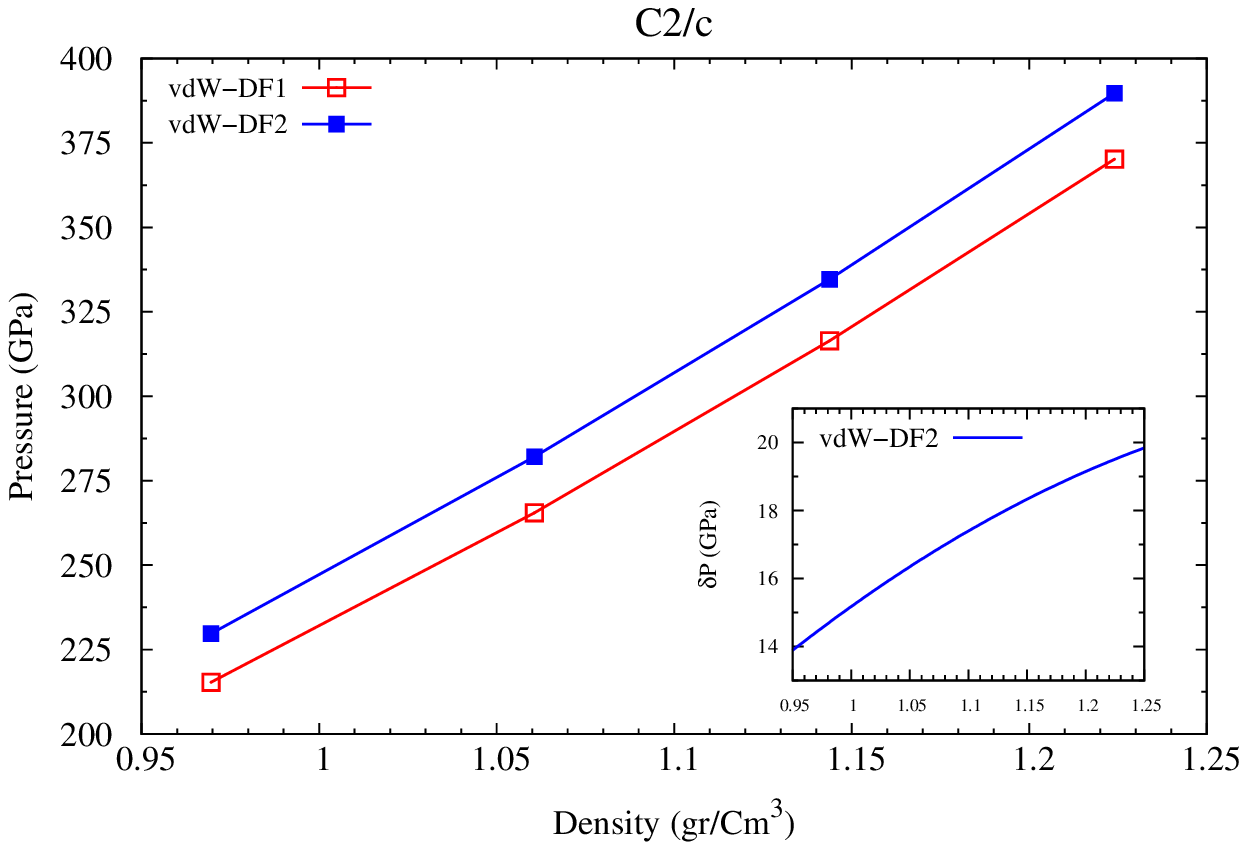}&
\includegraphics[width=0.45\textwidth]{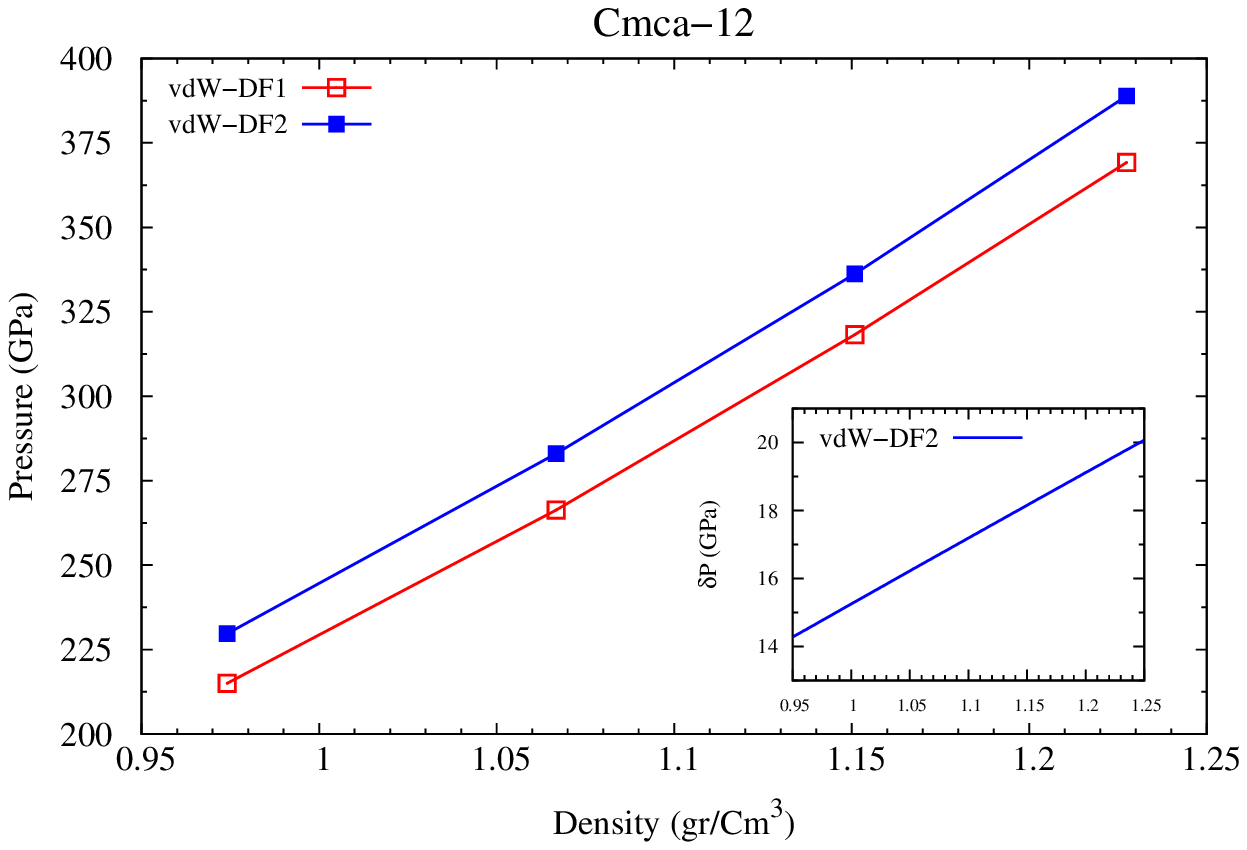}\\
\includegraphics[width=0.45\textwidth]{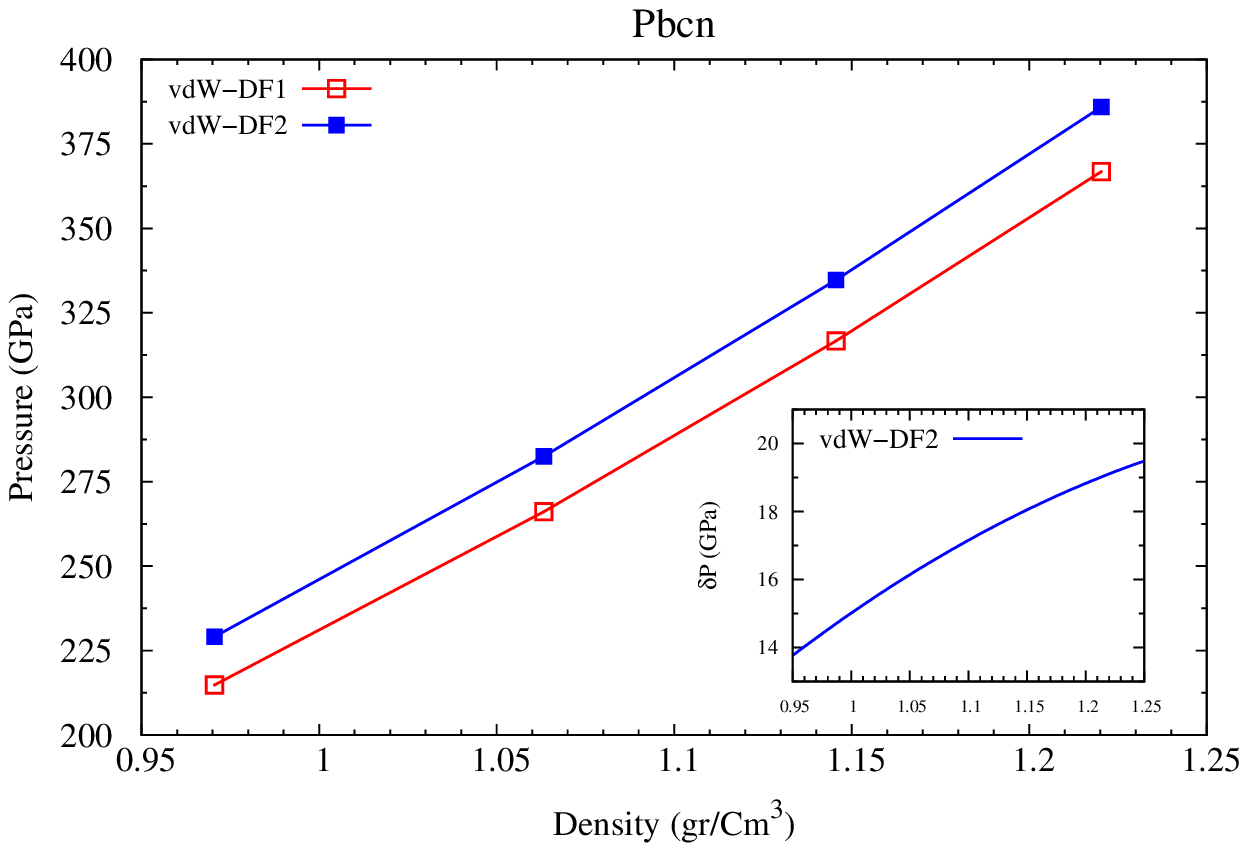}&
\includegraphics[width=0.45\textwidth]{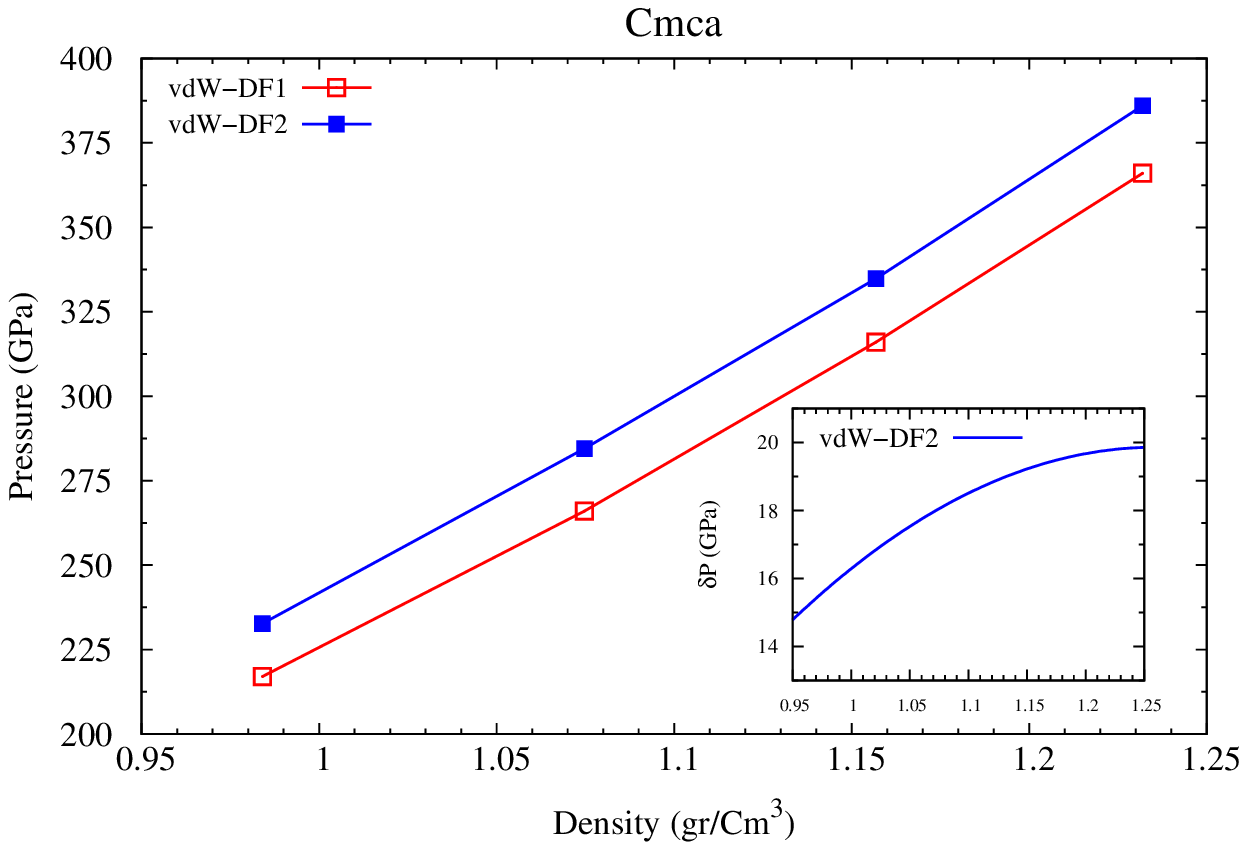}\\
\includegraphics[width=0.45\textwidth]{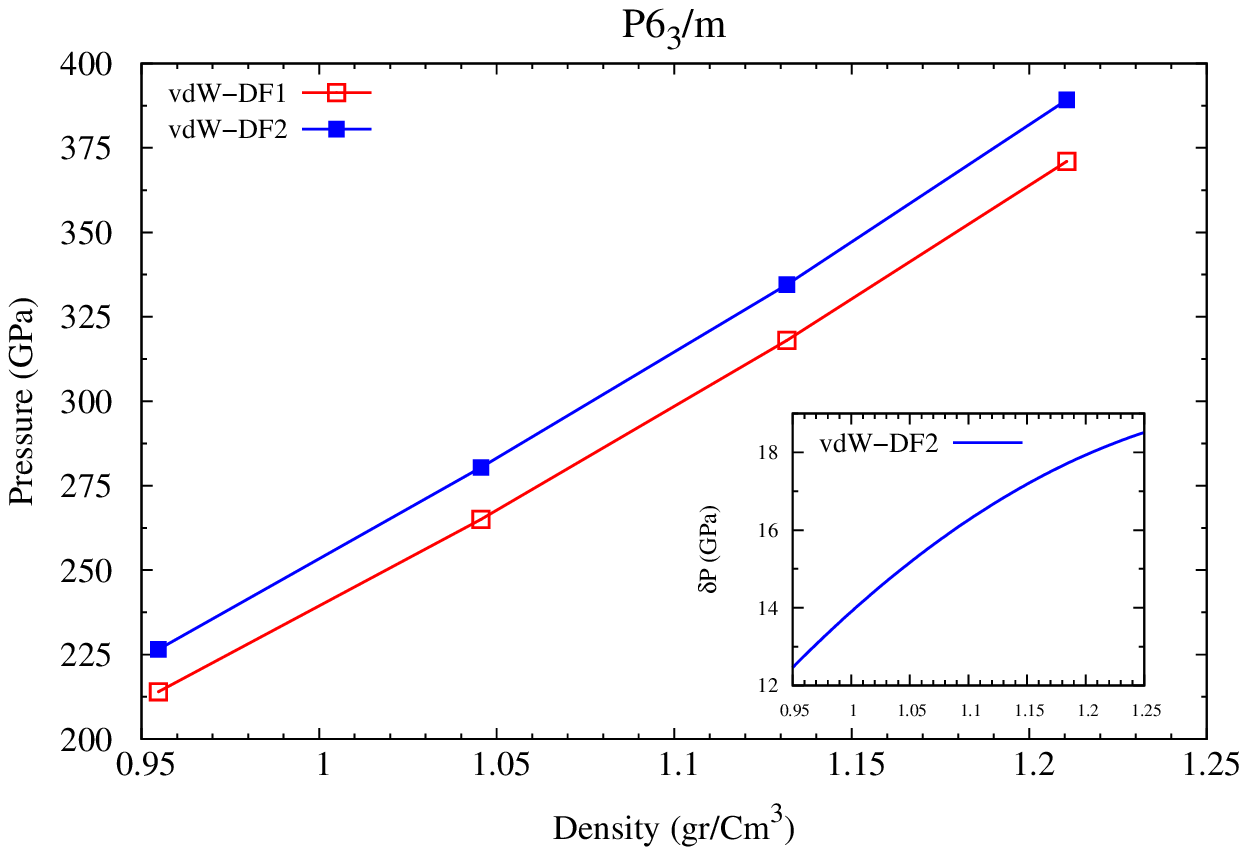}\\
\end{tabular}
\caption{\label{EOS1} (Color online) Pressure-density equation of state 
for the $C2/c$, $Cmca$-12, $Pbcn$, $Cmca$, and  $P6_3/m$
phases calculated using vdW-DF1 and vdW-DF2 functionals.
The inset shows evolution of $\delta$P,
the difference between vdW-DF2 pressure and vdW-DF1 pressure, as a function of density.
At a fixed density, vdW-DF2 predicts larger pressure than vdW-DF1.} 
\end{figure}
Figure\ref{EOS1} illustrates pressure-density equation of state for
the $C2/c$, $Cmca$-12, $Pbcn$, $Cmca$, and $P6_3/m$ molecular
structures which are obtained by vdW-DF1 and vdW-DF2 functionals.
Insets show the difference between pressure calcation at same density
($\delta P$), which increases with density and for all the studied
structures.  $\delta P$ is in the range 10-20GPa, and vdW-DF2 gives
systematically larger pressures than vdW-DF1.
\subsection{Lattice dynamics and Bondlengths}
\begin{figure}
\begin{tabular}{c c}
\includegraphics[width=0.4\textwidth]{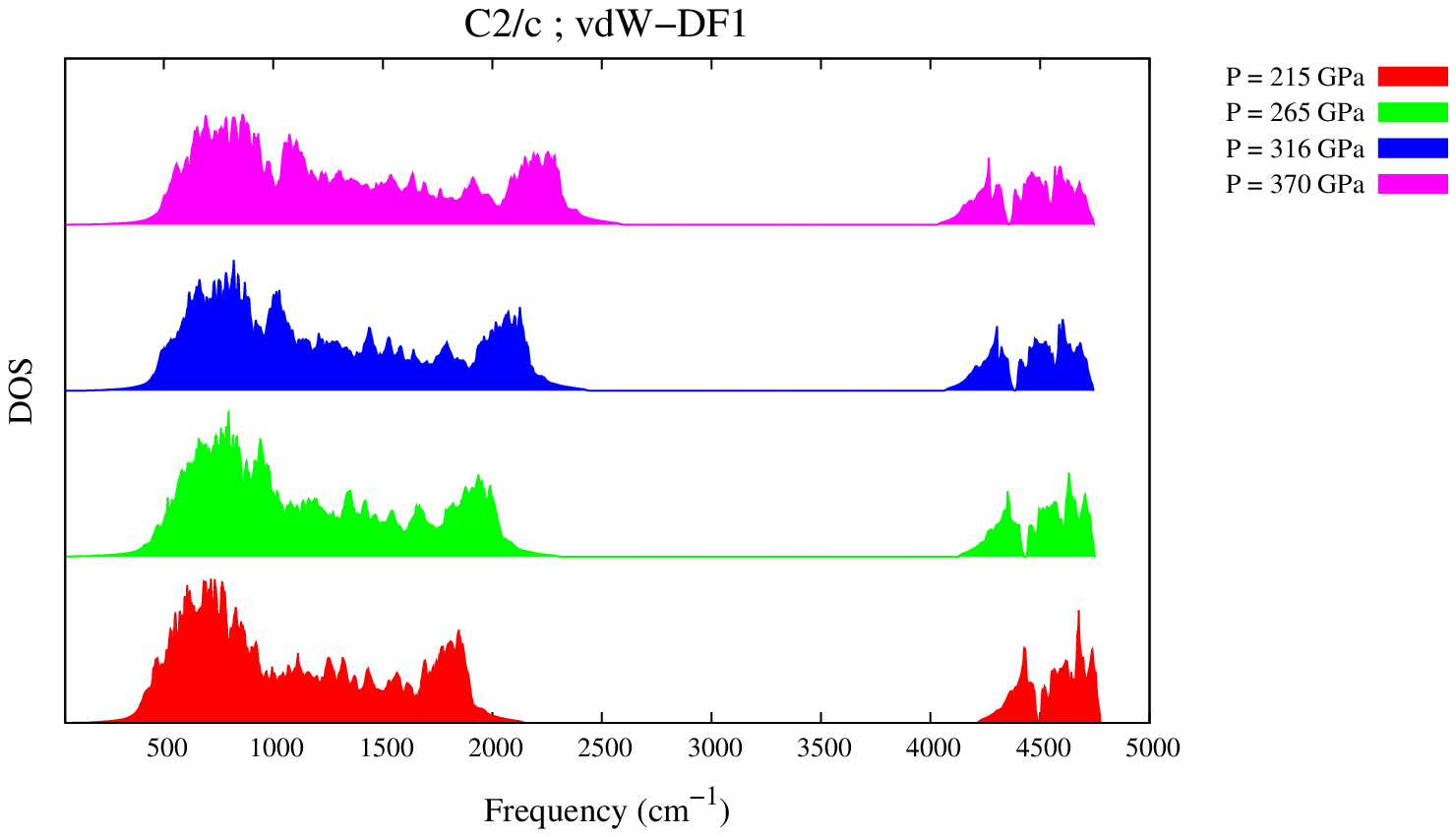}&
\includegraphics[width=0.4\textwidth]{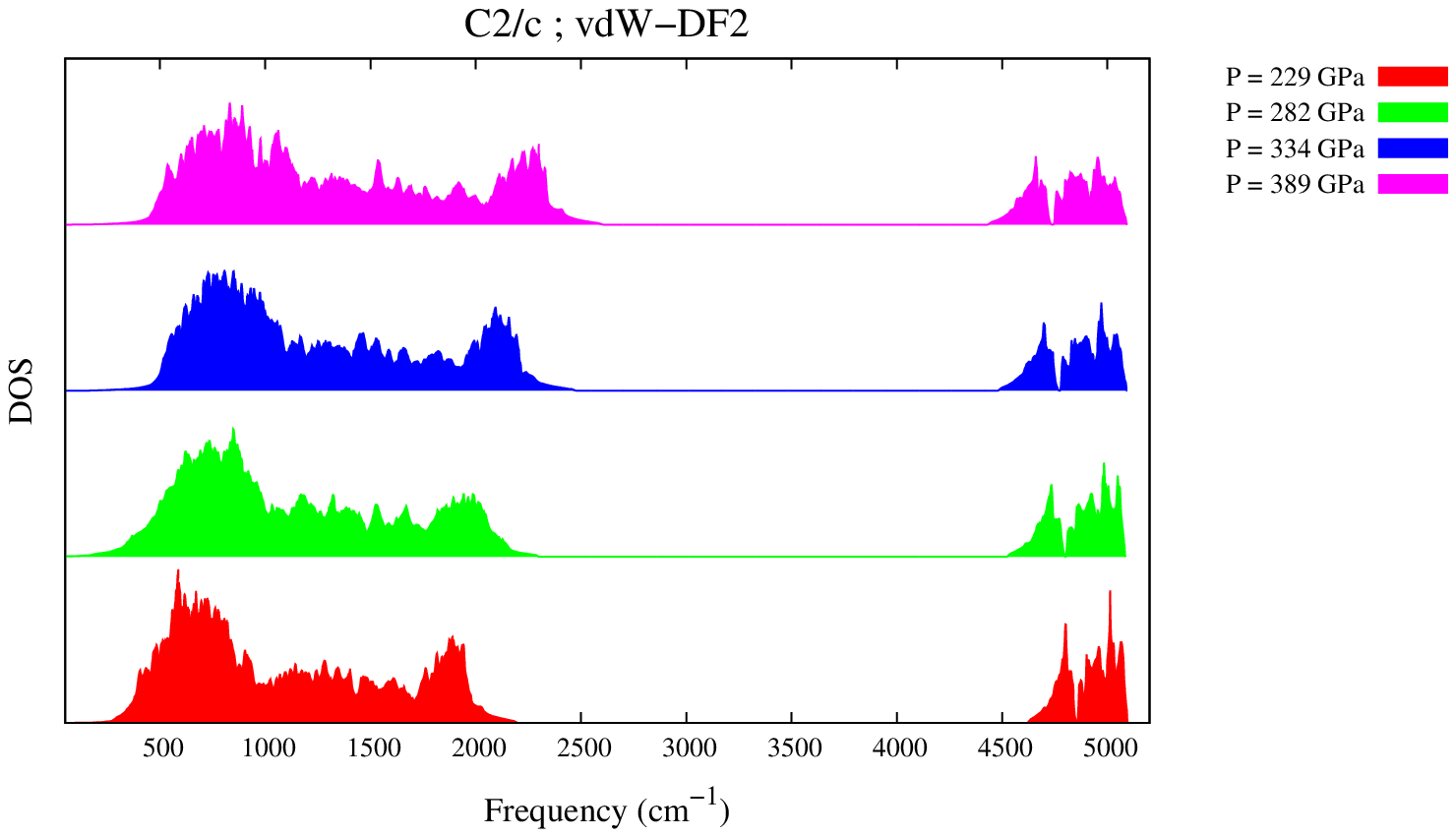}\\
\includegraphics[width=0.4\textwidth]{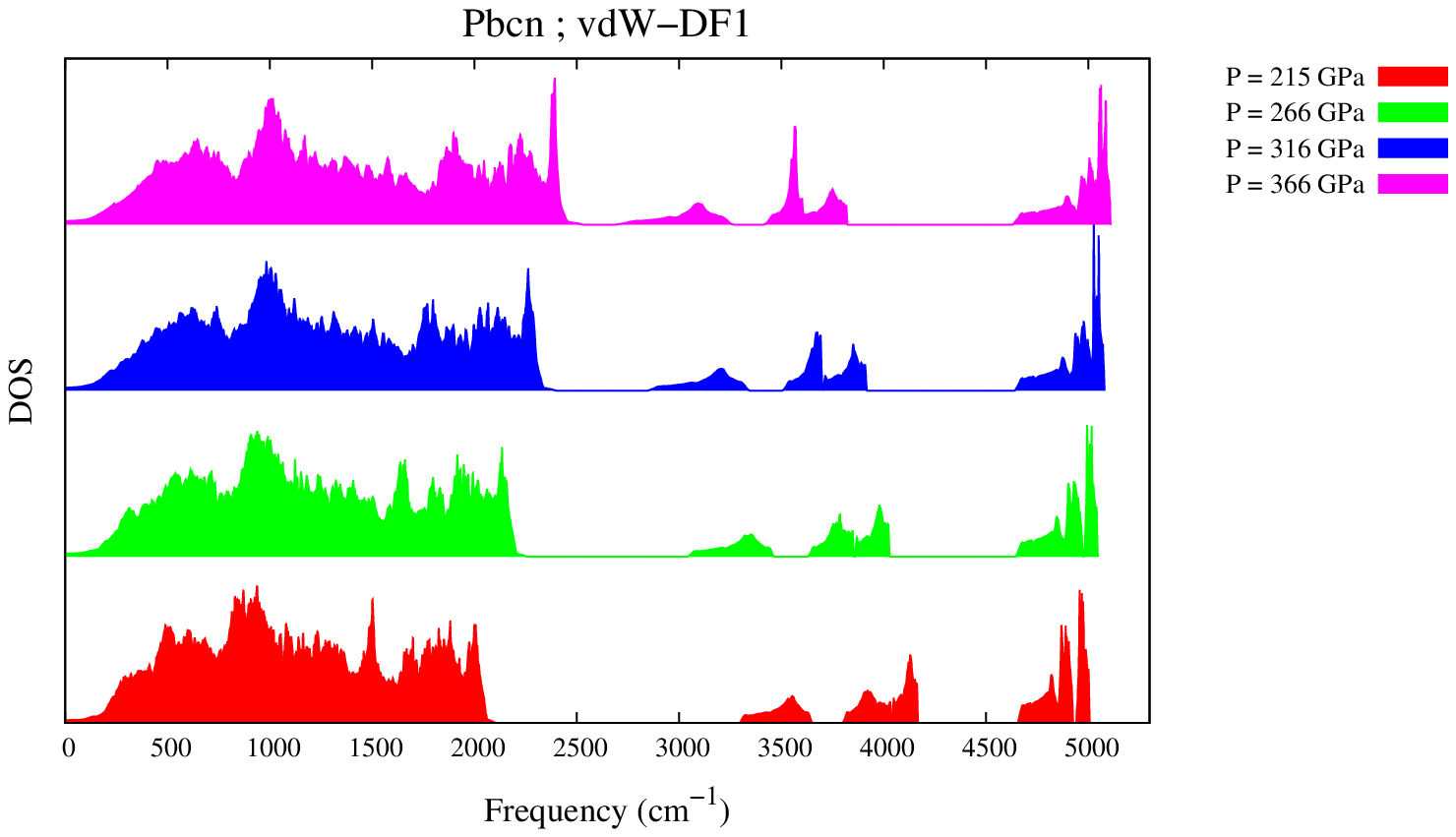}&
\includegraphics[width=0.4\textwidth]{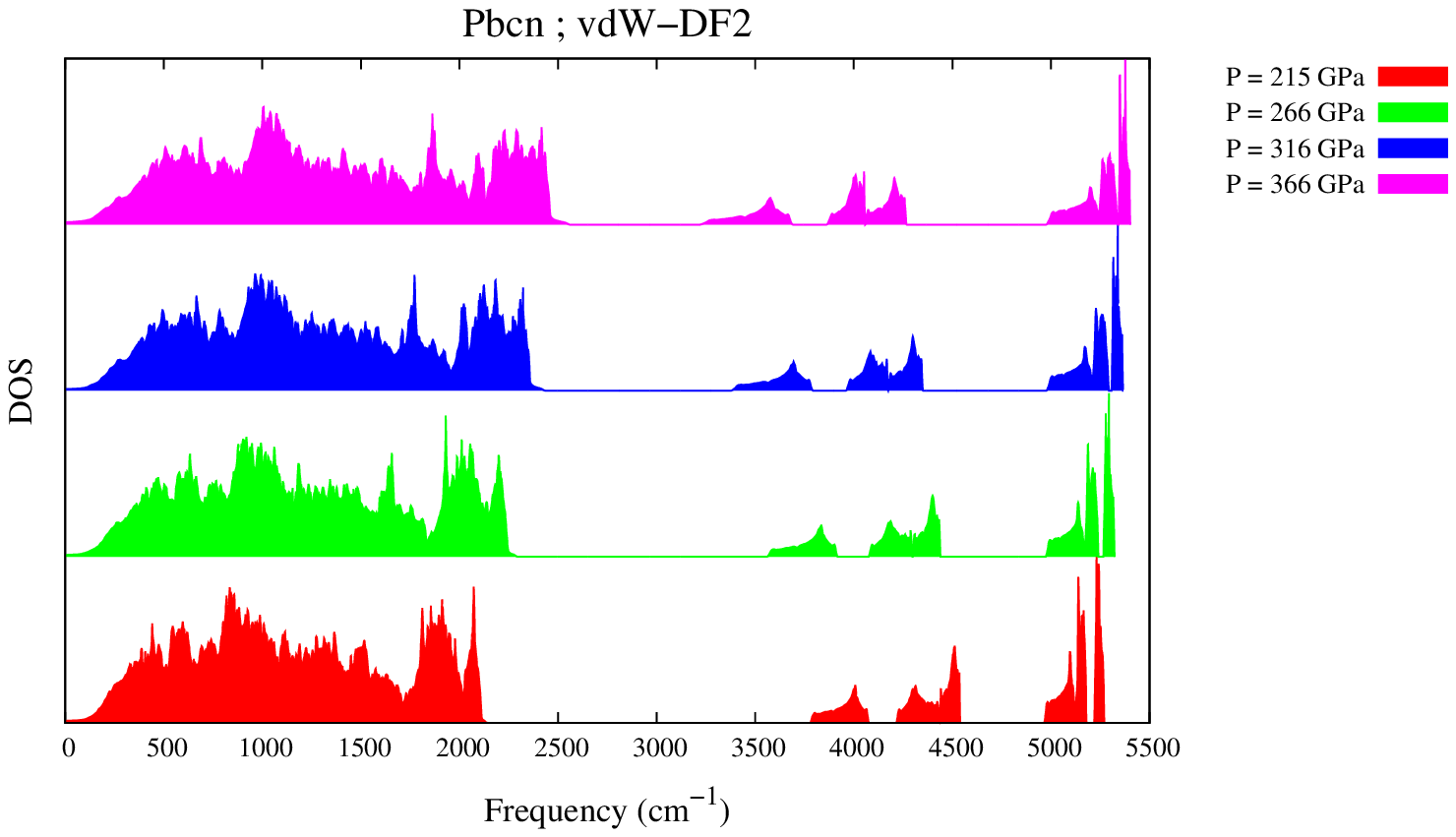}\\
\includegraphics[width=0.4\textwidth]{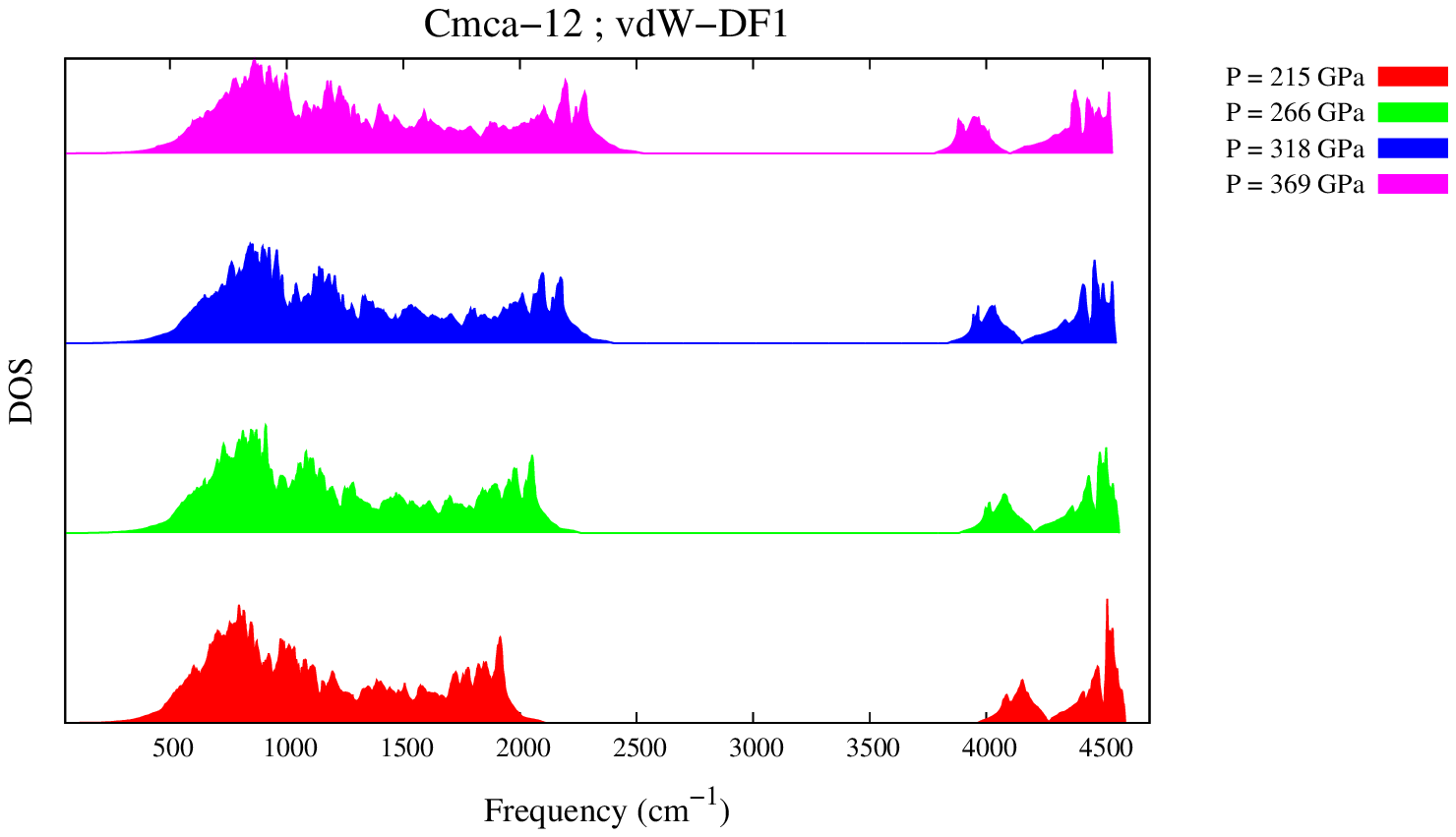}&
\includegraphics[width=0.4\textwidth]{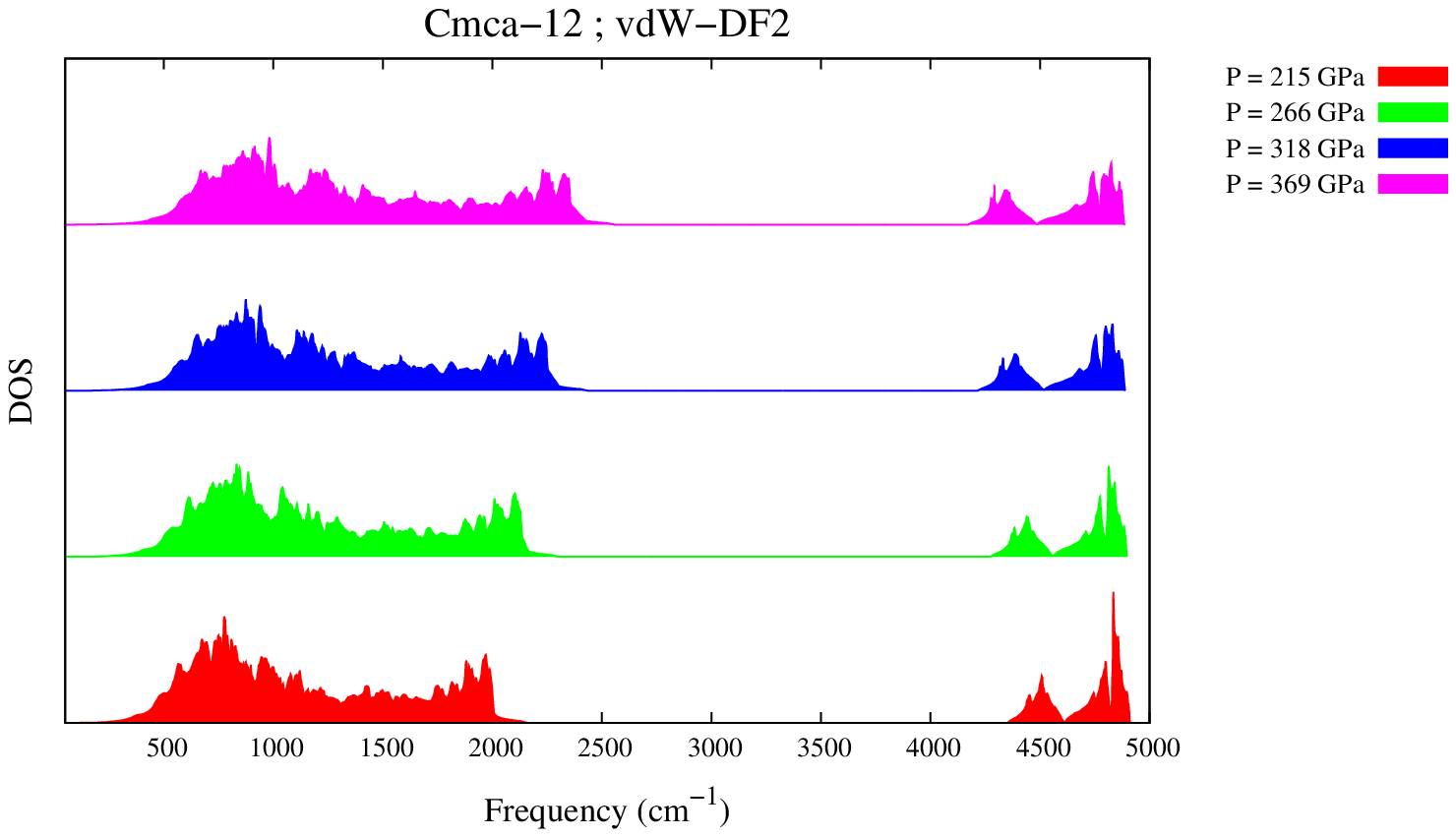}\\
\includegraphics[width=0.4\textwidth]{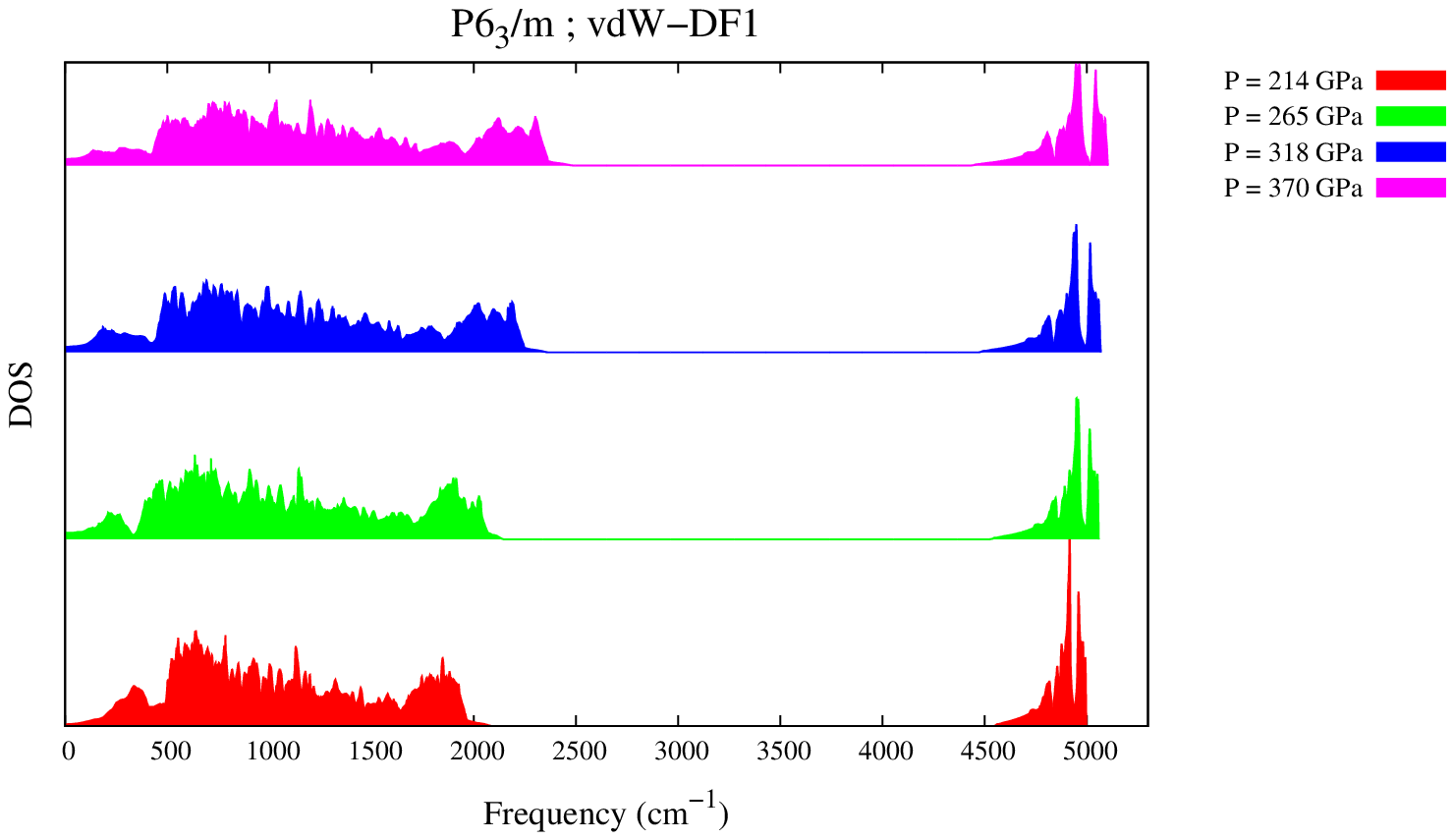}&
\includegraphics[width=0.4\textwidth]{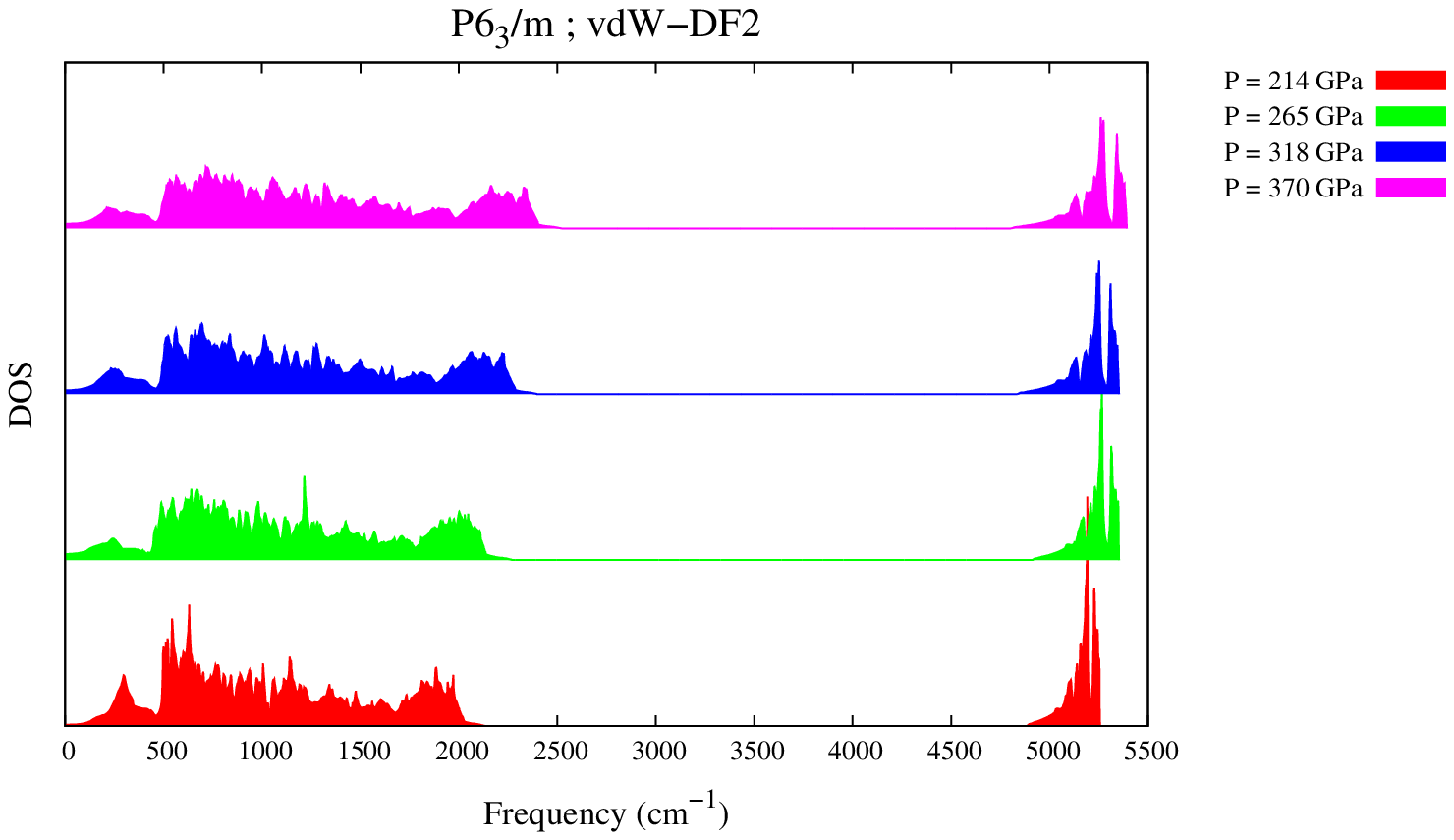}\\
\includegraphics[width=0.4\textwidth]{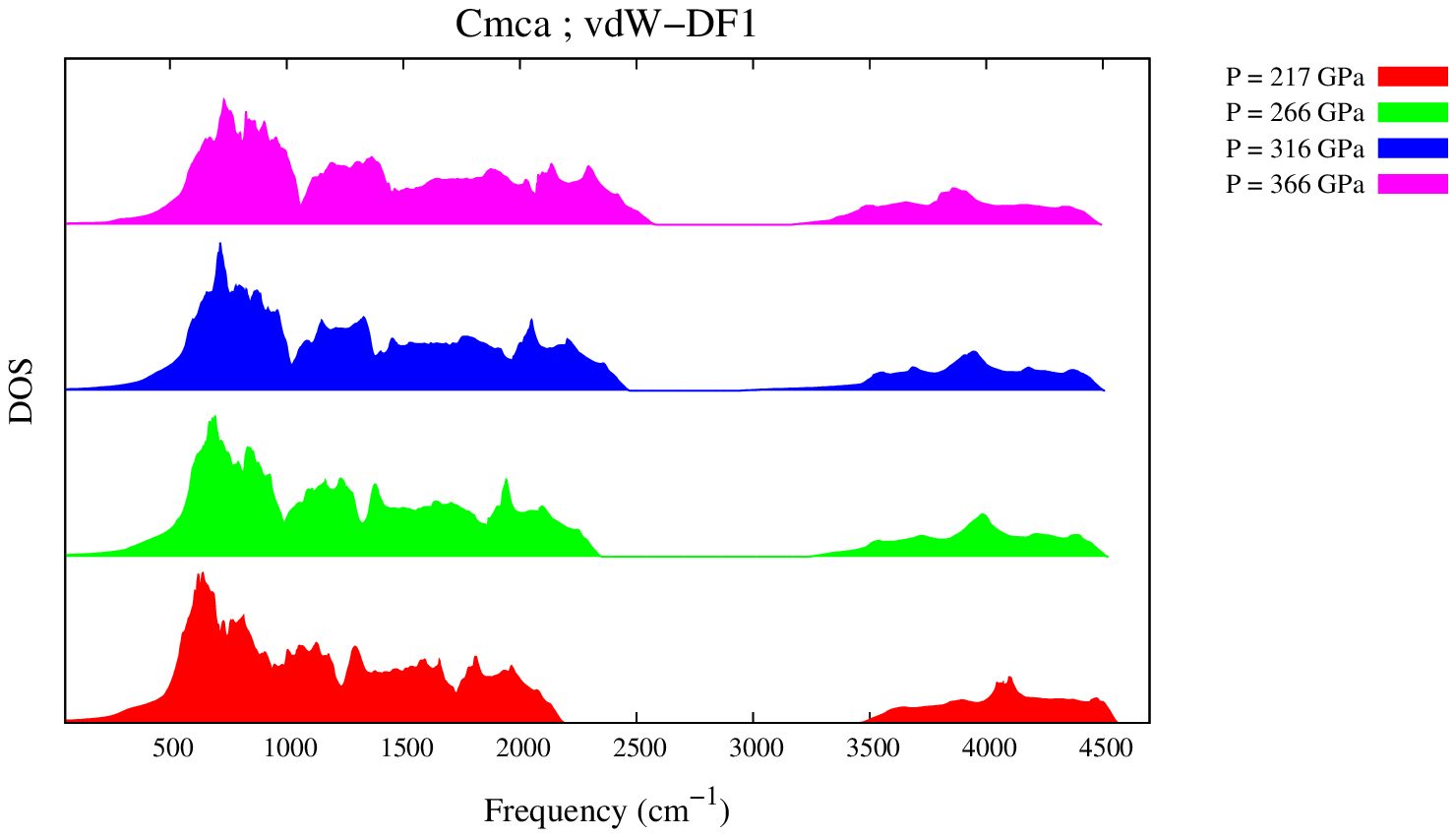}&
\includegraphics[width=0.4\textwidth]{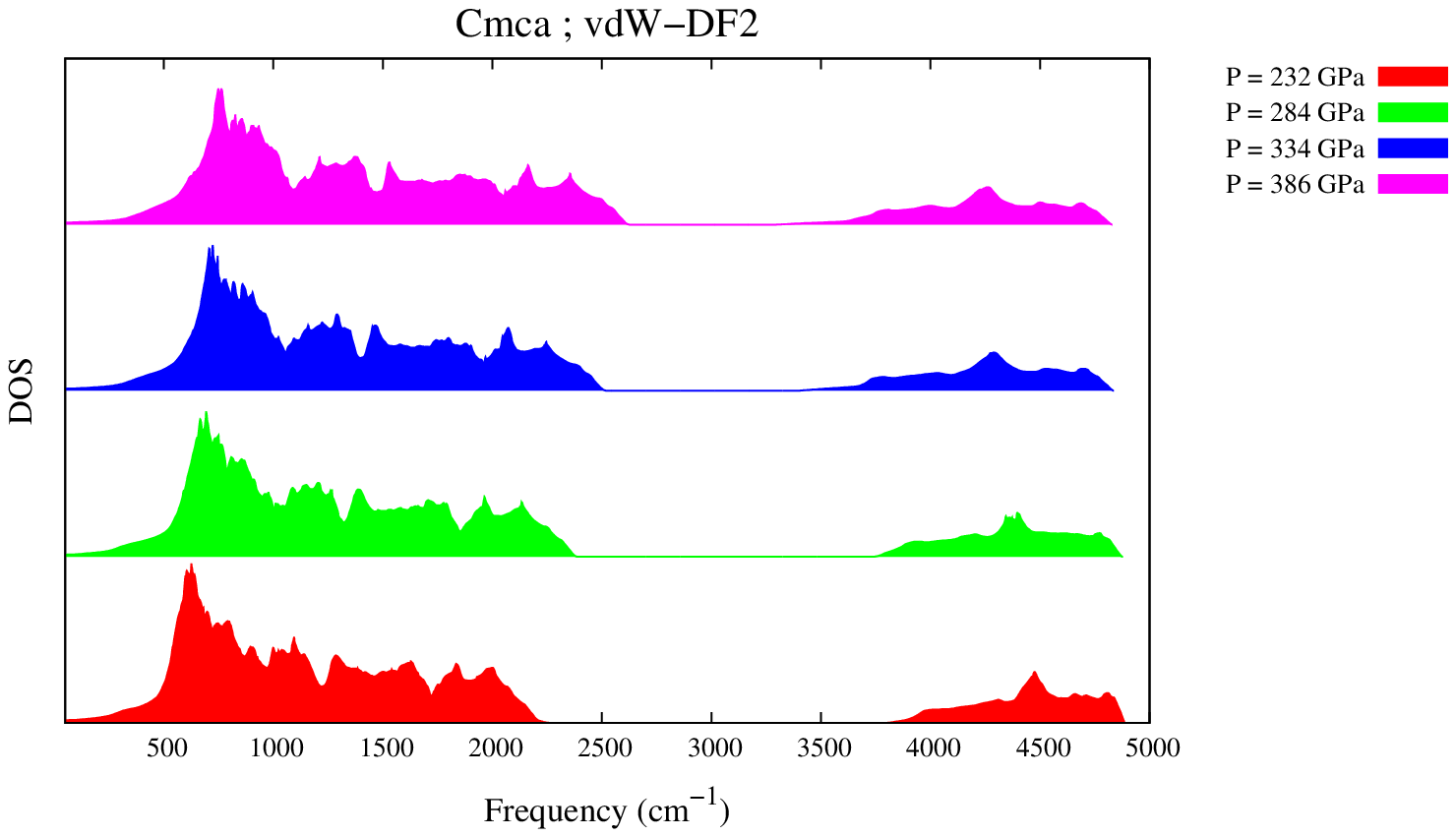}\\
\end{tabular}
\caption{\label{phonDOS} (Color online) The phonon density of states 
of the $C2/c$, $Pbcn$, $Cmca$-12, $P6_3/m$, and $Cmca$ 
phases calculated using vdW-DF1 and vdW-DF2 functionals at four 
different pressures. } 
\end{figure}

Figure \ref{phonDOS} illustrates phonon density of states (DOS) of 
the $C2/c$, $Pbcn$, $Cmca$-12, $P6_3/m$, and $Cmca$ 
structures which are obtained by vdW-DF1 and vdW-DF2 at four 
different pressures. Both vdW functionals predict that 
in all the studied molecular structures 
the phonon dispersion increases by increasing the  pressure.
The vibron frequencies predicted by vdW-DF1 are smaller than 
vdW-DF2. This difference is strongly related to the optimized $H-H$
molecular bond-length (BL) predicted by vdW functionals. 
Precise values of optimized molecular bond-length for 
all the studied structures, which are calculated by vdW-DF1 
and vdW-DF2 at same density, are presented in table \ref{TAB1}, respectively. 

\begin{figure}
\centering
\includegraphics[width=0.75\textwidth]{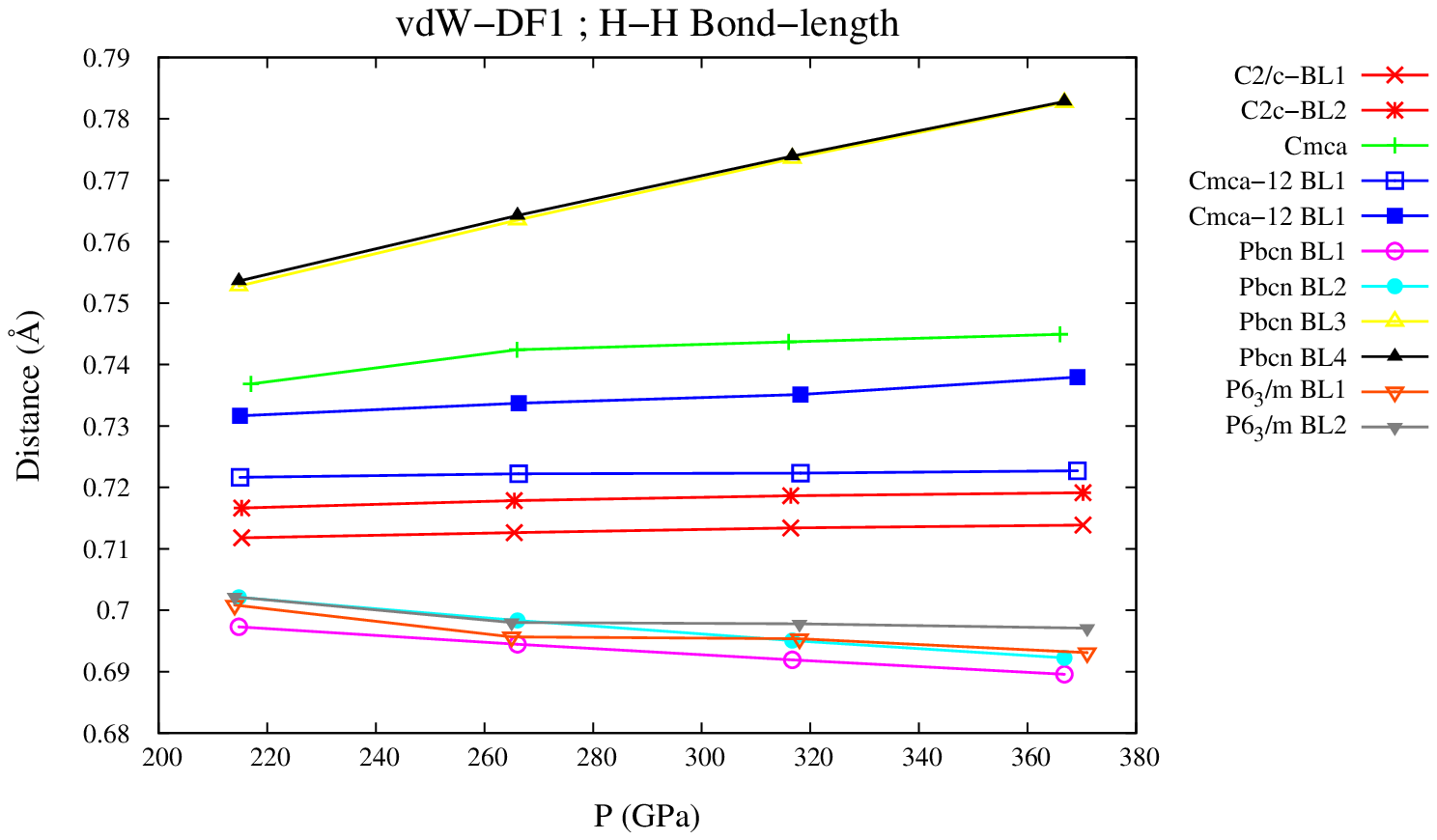}
\includegraphics[width=0.75\textwidth]{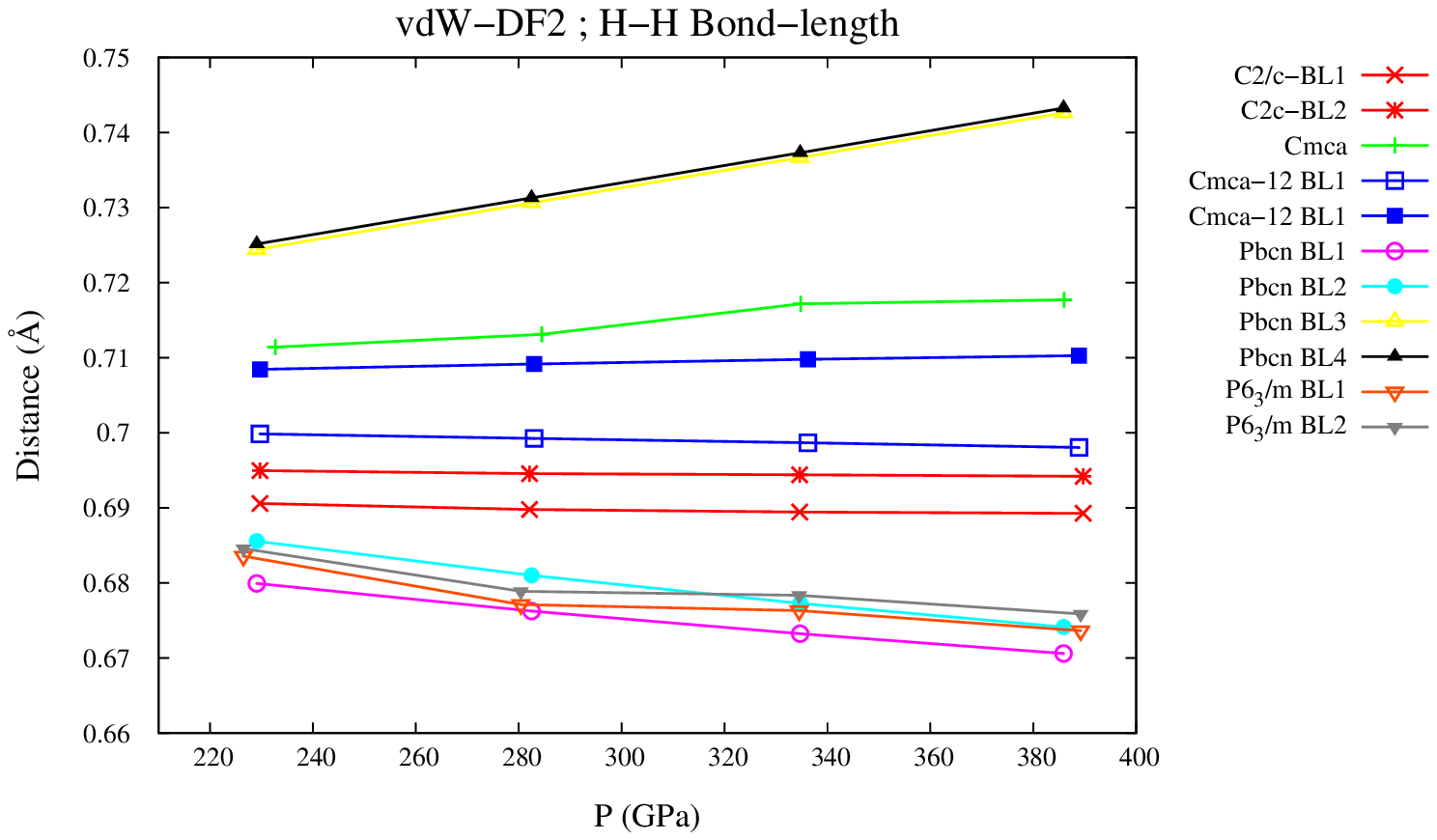}
\caption{\label{HHBL} (Color online) The optimized $H-H$ molecular bond-length (BL) 
of the $C2/c$, $Pbcn$, $Cmca$-12, $P6_3/m$, and $Cmca$ 
phases calculated using vdW-DF1 and vdW-DF2 functionals at
different pressures. In the primitive unit cells of the $C2/c$, $Pbcn$, $Cmca$-12, $P6_3/m$, and $Cmca$
structures there are two, four, two, two, and one categories of $H_2$ molecules 
with different $H-H$ bond-lengths.} 
\end{figure}

In the primitive unit cell of $C2/c$ there are twelve $H_2$ molecules
with two kinds of $H-H$ molecular bond-length  named BL1 and BL2
in figure \ref{HHBL}. The $C2/c-$BL1 which is shorter than $C2/c-$BL2
corresponds to higher vibron frequencies.
vdW-DF1 and vdW-DF2 produce different results for the $C2/c-$BL1 and
$C2/c-$BL2.  According to vdW-DF1 results both $C2/c-$BL1 and $C2/c-$BL2
increase with density, whereas vdW-DF2 predict that $C2/c-$BL1 and
$C2/c-$BL2 slightly decrease by increasing the pressure (Table
\ref{TAB1}).  The $C2/c-$BL1 and $C2/c-$BL2 obtained by
vdW-DF1 are larger than those calculated by vdW-DF2 at the same
density. Therefore, vdW-DF2 functional predict higher frequency
vibrons. Same argument can be applied on the $Cmca-12$ phase.  The
difference between $Cmca$-12$-$BL1 and $Cmca$-12$-$BL2 is larger than
the difference between $C2/c-$BL1 and $C2/c-$BL2. 
Hence, the $Cmca$-12 phonon vdW's DOS
results predict lower frequencies for vibrons than $C2/c$.
$Cmca$-12$-$BL1 and $Cmca$-12$-$BL2 obtained by vdW-DF1 both increase
by density.  But vdW-DF2 optimized molecular bond-length indicate that
$Cmca$-12$-$BL1 and $Cmca$-12$-$BL2 decreases and increases by
pressure, respectively.  All the $H-H$ molecular bond-lengths in the
$Cmca$ phase are identical and consequently one vibron can be observed
in phonon DOS (Figure \ref{HHBL}).  The vdW-DF1 and vdW-DF2 predict
that $Cmca$-BL increases by rising pressure.  $Cmca$-BL is larger than
$C2/c$ and $Cmca$-12 molecular bond-lengths and therefore the vibron
frequency of the $Cmca$ is smaller than $C2/c$ and $Cmca$-12
structures.

Recently we have used many-body wave-function based 
quantum Monte Carlo methods to calculate excitonic and quasi-particle band gap 
and also electronic band-structure of the $C2/c$, $Pbcn$, and $P6_3/m$ phases\cite{PRB17}.
We have discovered that many properties of solid molecular hydrogen are strongly correlated
with $H-H$ molecular bond-length. For instance, the
gradient of the $P6_3/m$ band gap with respect to molecular BL
is $\sim27.3$ eV$/\AA$ independent of the XC functional. 
As illustrated in Figure \ref{HHBL}, the twenty four $H_2$ molecules
in the unit cell of the $Pbcn$ phase adopt four non-equivalent $H-H$
molecular bond-lengths, and thus four vibron frequencies are
obtained. $Pbcn-$BL1 and $Pbcn-$BL2, which shorten with pressure
are shorter than $Pbcn-$BL3 and $Pbcn-$BL4,
which lengthen with pressure(Table
\ref{TAB1}).  This behaviour correlates with
the $Pbcn$ phonon DOS where increasing the pressure 
reduces the two low vibron frequencies and increases two high vibron
frequencies. 
The $P6_3/m$ structure 
has eight $H_2$ molecules per primitive unit cell with two inequivalent
bond-lengths $P6_3/m-$BL1 and $P6_3/m-$BL2 with
difference of $\sim 4m\AA$. VdW-DF1 and vdW-DF2 results indicate that
$P6_3/m-$BL1 and $P6_3/m-$BL2 are reduced by increasing density.

The averaged molecular bond-length by vdW-DF1, 
which is defined as $BL_{ave}=1/n\sum_i^nBL_i$ where n is the number of bond-lengths, 
of all the studied phases increase by pressure except the $P6_3/m$. 
vdW-DF2 calculations show that $BL_{ave}$ reduction of the $C2/c$ and $P6_3/m$
due to increasing the pressure are 1m$\AA$ and 9m$\AA$, respectively. 
Similar to vdW-DF1 results the $BL_{ave}$ of other structures calculated by vdW-DF2 
are increases by dense. By increasing
the pressure both low energy lattice phonon frequencies and high frequency vibron 
modes become larger. Decreasing the $P6_3/m$-BL$_{ave}$ with pressure causes an instability 
in the system which is also found from imaginary phonon frequencies. It should be 
noted that at high enough pressures molecular to atomic phase transition occurs.  
The Raman spectra and MD simulations suggest that phase IV is a mixture 
elongated $H_2$ dimers experiencing large pairing fluctuations,
and unbound $H_2$ molecules\cite{Howie}. This matches very well with the altering of $Pbcn$ 
molecular bond-lengths with pressure. The $Pbcn-$BL1 and $Pbcn-$BL2 decrease with increasing the 
pressure whereas $Pbcn-$BL4 and $Pbcn-$BL3 dissociate by increasing the pressure. 
Benchmarking DFT-XC functionals for high pressure solid hydrogen using quantum
Monte Carlo (QMC) simulations indicate that, at static level,
optimized molecular bond-length for the $C2/c$ which is calculated by
vdW-DF1 functional agrees with QMC results\cite{Clay}.  

\begin{table}[h!]
\centering
\begin{tabular}{|c|cc|c|cc|cccc|cc|} 
\specialrule{.3em}{.2em}{.2em} 
vdW-DF1 & $C2/c$ & & $Cmca$ & $Cmca$-12& & & $Pbcn$ & &  &$P6_3/m$& \\
\hline
 P & BL1   & BL2 & BL    & BL1     &BL2 & BL1         & BL2 & BL3 & BL4 & BL1    & BL2 \\
\hline
215& 0.71179& 0.71664& 0.73683& 0.72164& 0.73165& 0.69731& 0.70211& 0.75279& 0.75361& 0.70085& 0.70219 \\
265& 0.71265& 0.71784& 0.74240& 0.72222& 0.73371& 0.69447& 0.69833& 0.76352& 0.76427& 0.69565& 0.69802 \\
316& 0.71341& 0.71865& 0.74370& 0.72233& 0.73511& 0.69193& 0.69505& 0.77351& 0.77394& 0.69537& 0.69780 \\
370& 0.71386& 0.71912& 0.74491& 0.72272& 0.73795& 0.68958& 0.69224& 0.78263& 0.78283& 0.69309& 0.69707 \\
\specialrule{.3em}{.2em}{.2em} 
\hline
vdW-DF2 &&&&&&&&&&&\\
230& 0.69058& 0.69496& 0.71141& 0.69985& 0.70846& 0.67994& 0.68557& 0.72442& 0.72514& 0.68358& 0.68459 \\ 
282& 0.68979& 0.69457& 0.71311& 0.69924& 0.70916& 0.67625& 0.68102& 0.73063& 0.73129& 0.67712& 0.67888 \\
335& 0.68945& 0.69440& 0.71719& 0.69868& 0.70980& 0.67324& 0.67727& 0.73666& 0.73730& 0.67633& 0.67835 \\
390& 0.68927& 0.69421& 0.71771& 0.69804& 0.71029& 0.67061& 0.67413& 0.74261& 0.74325& 0.67364& 0.67587 \\
\specialrule{.3em}{.2em}{.2em} 
\end{tabular}
\caption{\label{TAB1} Precise values of molecular bond-length in $\AA$
  for the $C2/c$, $Cmca$, $Cmca$-12, $Pbcn$, and $P6_3/m$ structures
  calculated by vdW-DF1 and vdW-DF2 at various pressures (in GPa).  
First column shows the pressure (P) in GPa. Notice how the
  shorter bond lengths become shorter with pressure, while longer BLs
  become longer.}
\end{table}

The molecular bond-length, which is also the nearest-neighbour distance, is strongly 
correlated with electronic energy band gap. We do not present DFT band gap results here as our recent
extensive study of energy band gap of solid molecular hydrogen is reported in reference
\onlinecite{PRB17}. At constant pressure, the $P6_3/m$ band gap is larger than other 
molecular phases. As it is shown in figure \ref{HHBL}, the $P6_3/m-$BL$_{ave}$ is smaller 
than other studied structures. Shortening molecular bond-length localises the electrons and 
increase localised charge density and consequently, according to the band theory, opening the 
energy band gap is expected.  Hence, a precise bond length is necessary for an accurate
prediction of the properties of solid molecular phases.  
We propose a rule of thumb of \textit{the shorter molecular bond-length
the larger electronic band gap the higher vibron frequencies} which is independent
of the XC functional in DFT calculations study of high-pressure solid molecular hydrogen.
We have recently calculated the \textit{scissor operator} for solid molecular
hydrogen structures and we have demonstrated that the \textit{scissor operator} is also 
independent of DFT XC functional\cite{PRB13}. 

\begin{figure}
\begin{tabular}{c c}
\includegraphics[width=0.45\textwidth]{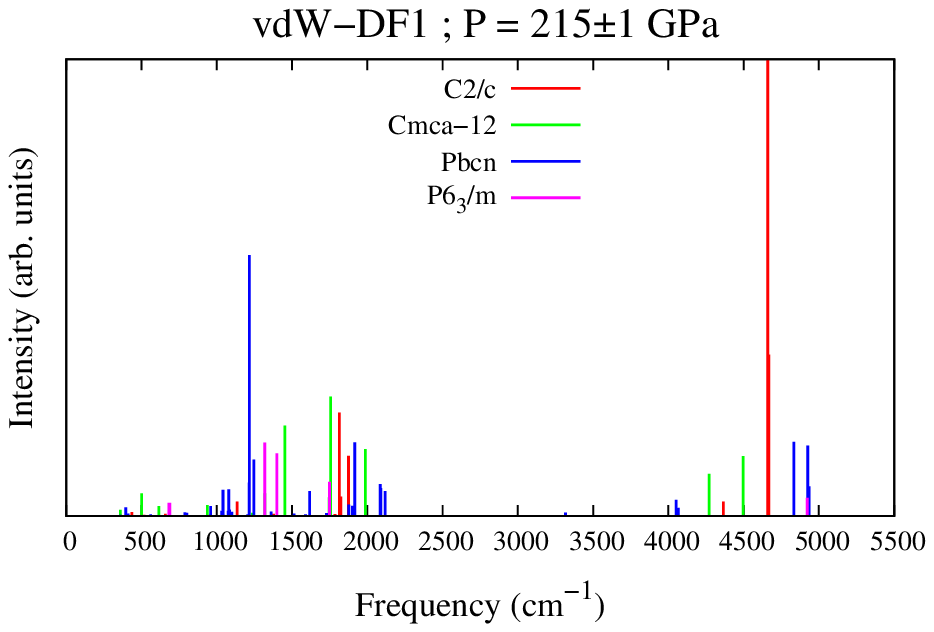}&
\includegraphics[width=0.45\textwidth]{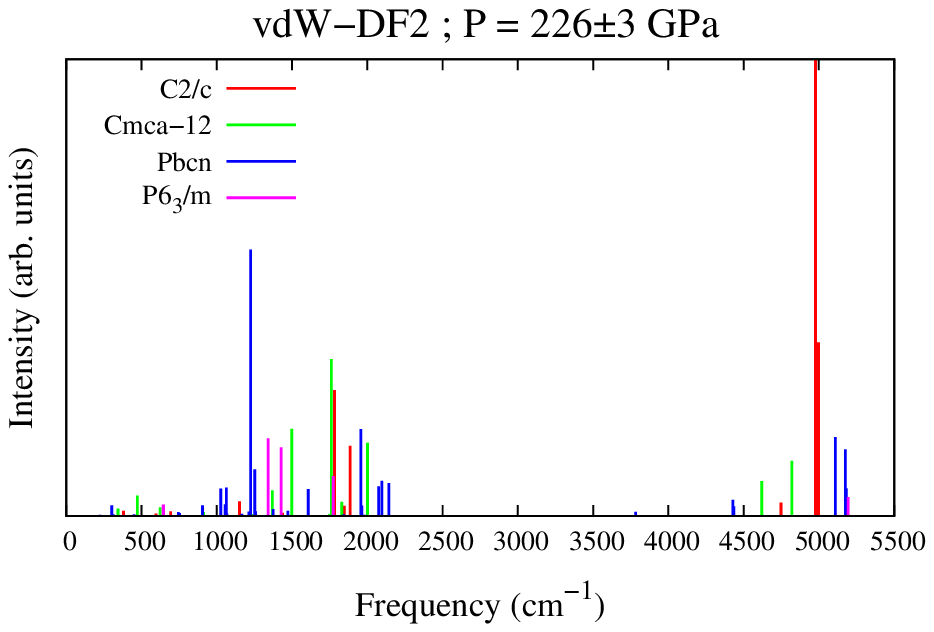}\\
\includegraphics[width=0.45\textwidth]{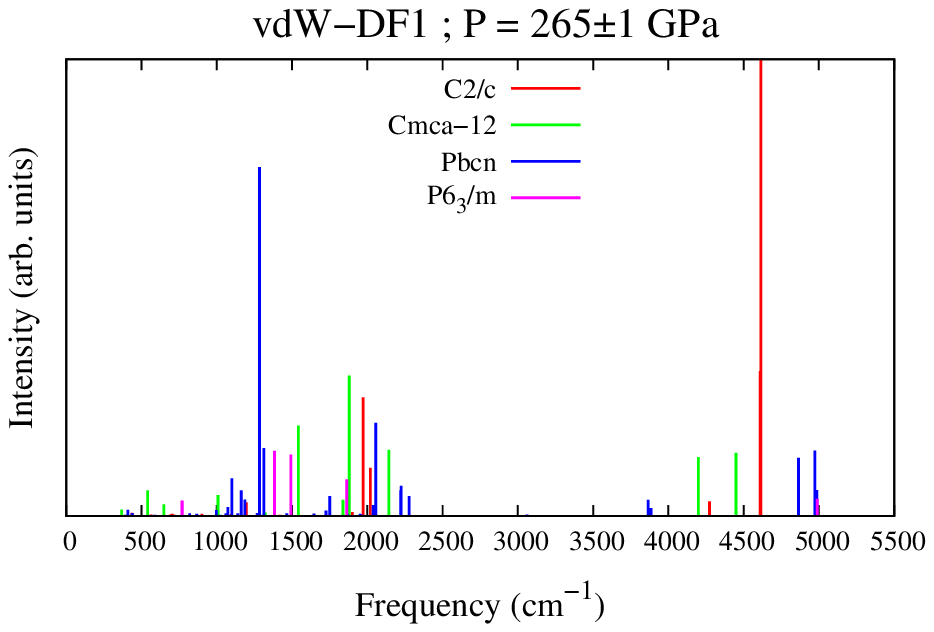}&
\includegraphics[width=0.45\textwidth]{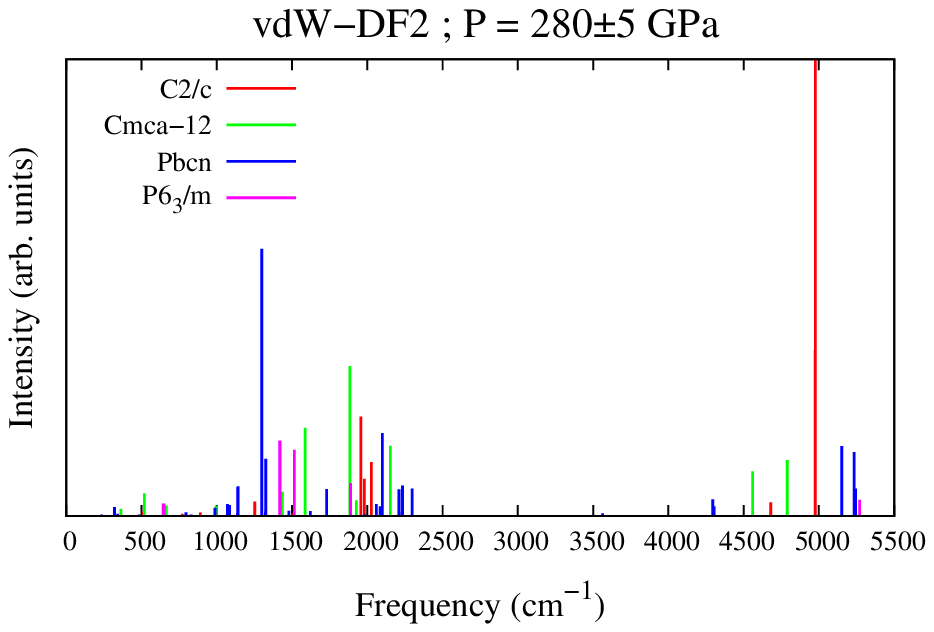}\\
\includegraphics[width=0.45\textwidth]{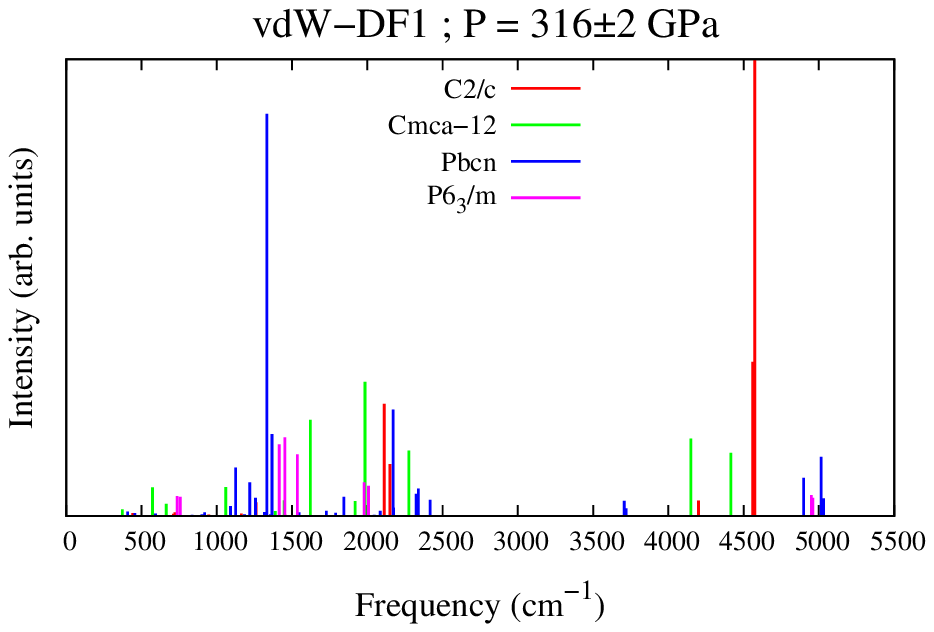}&
\includegraphics[width=0.45\textwidth]{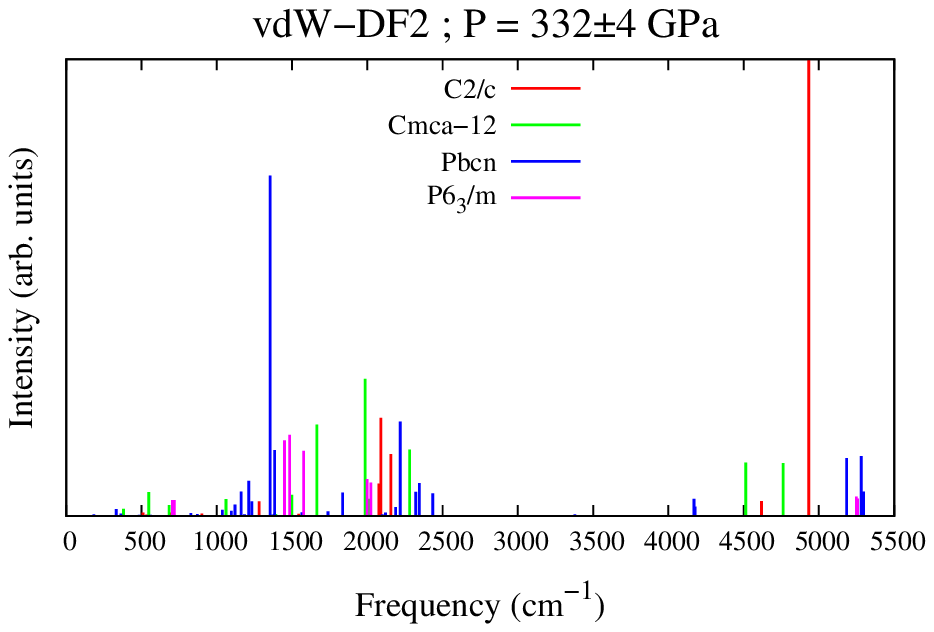}\\
\includegraphics[width=0.45\textwidth]{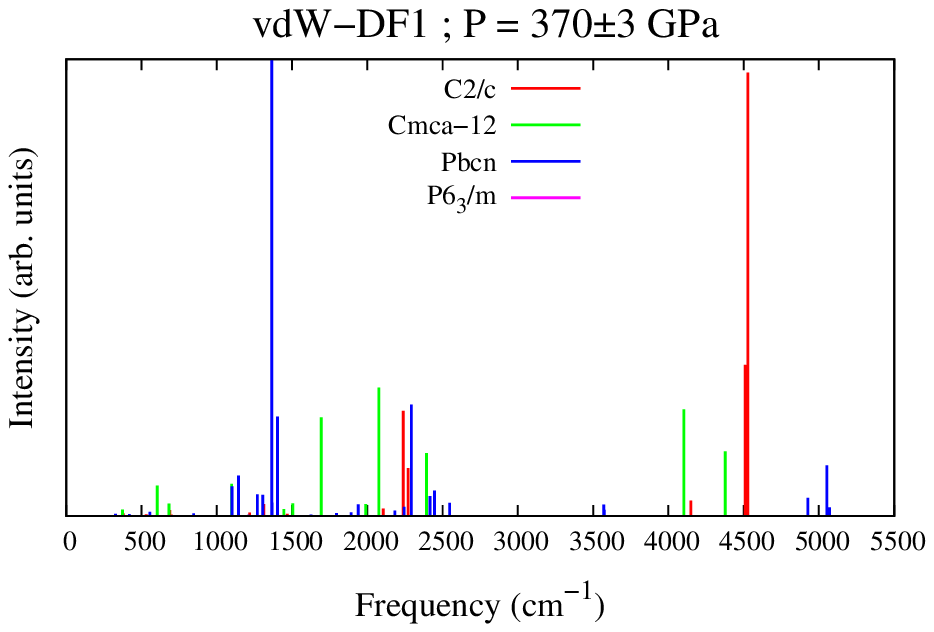}&
\includegraphics[width=0.45\textwidth]{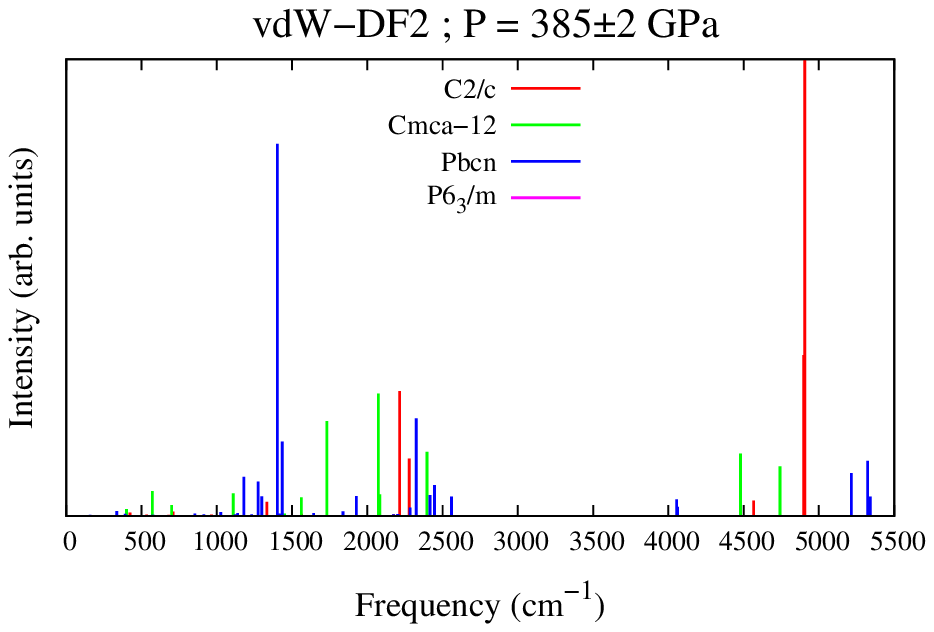}\\
\end{tabular}
\caption{\label{IR} (Color online) IR frequencies and relative intensities for
the $C2/c$, $Pbcn$, $Cmca$-12, $P6_3/m$, and $Cmca$ 
phases calculated using vdW-DF1 and vdW-DF2 functionals at four 
different pressures.} 
\end{figure}

We also calculated active IR modes for the solid molecular phases,
$C2/c$, $Pbcn$, $Cmca$-12, $P6_3/m$, and $Cmca$ structures which are
calculated by vdW-DF1 and vdW-DF2 at four different pressures ( Figure
\ref{IR} ).  The main difference between vdW IR spectra and those
simulated using conventional semi-local functionals is the position
of peak. The position of IR peaks depends on the optimized BL
predicted by XC functional. The gradient of IR peak with respect to BL
is $\sim 14.95\pm0.5 {(cm.m\AA)}^{-1}$ which is independent of
pressure and is also identical for the studied molecular
structures. It indicates that altering bond-length by $0.1 \AA$, which
equals to accuracy of DFT functionals in prediction of optimized BL
for $H_2$ molecule, shifts the IR peak by $1495\pm5 (cm)^{-1}$. This
value is almost same as the gap in phonon density of states.
\subsection{Finite temperature phase diagram}
\begin{figure}
\begin{tabular}{c c}
\includegraphics[width=0.5\textwidth]{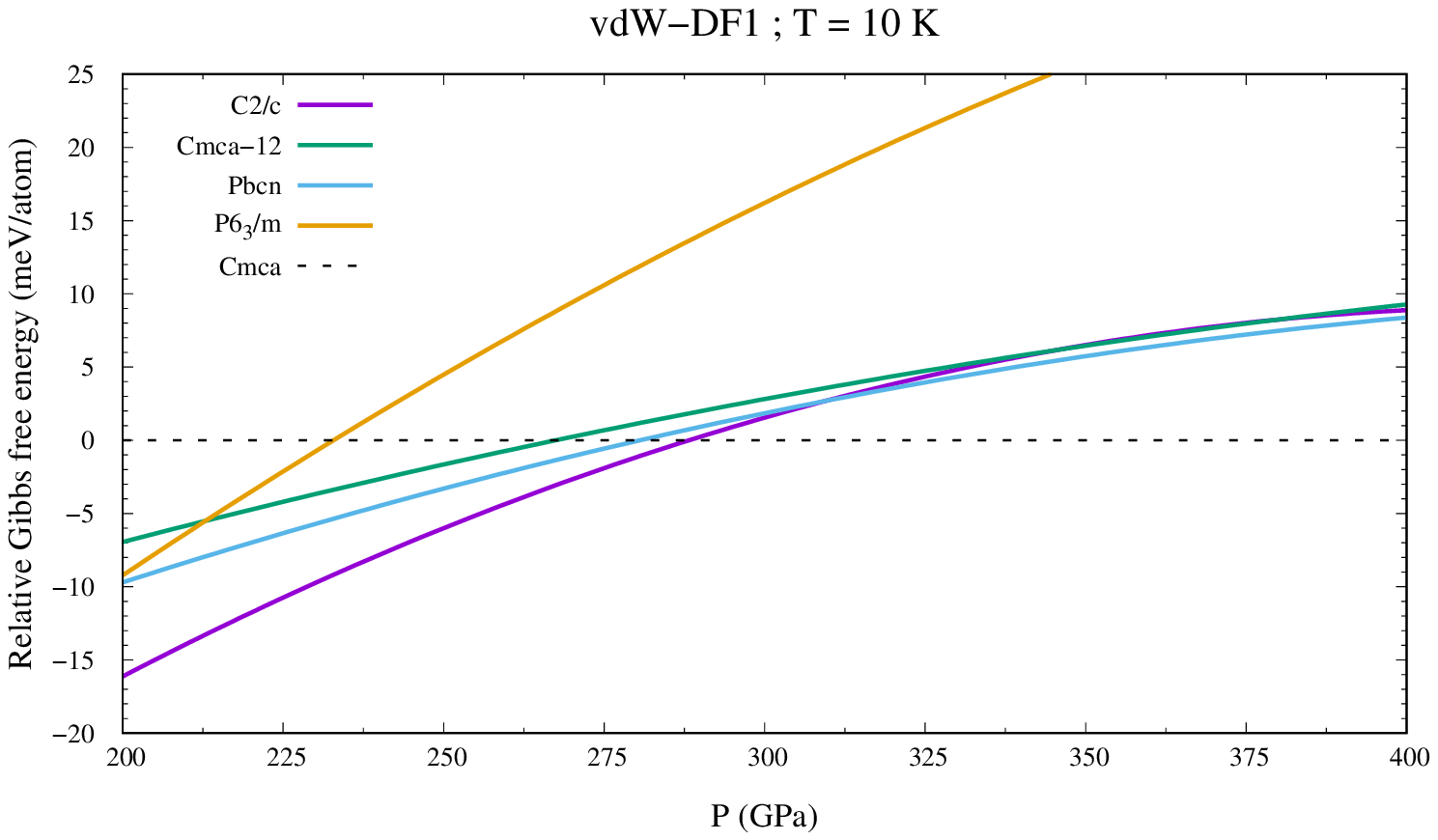}&
\includegraphics[width=0.5\textwidth]{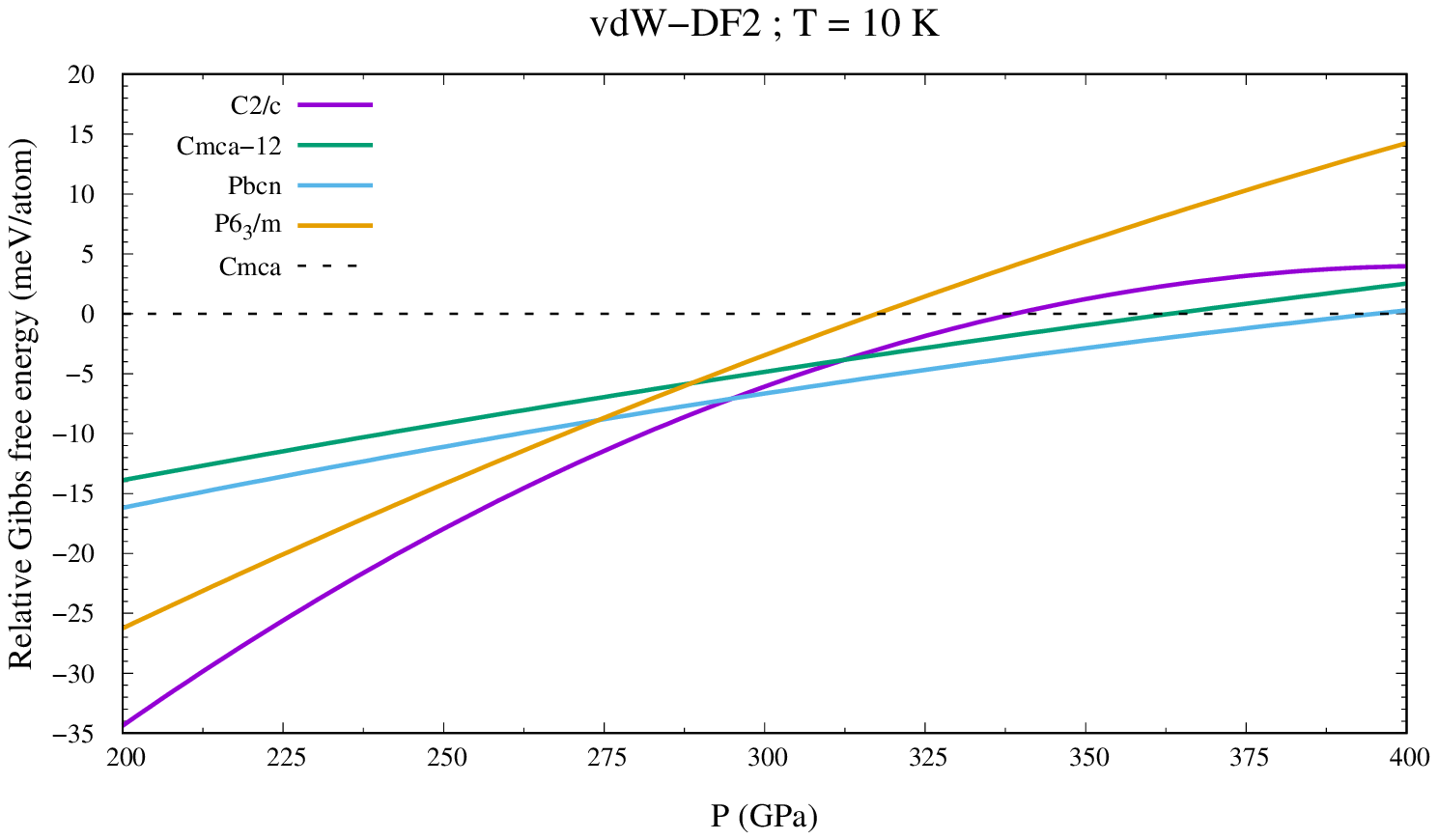}\\
\includegraphics[width=0.5\textwidth]{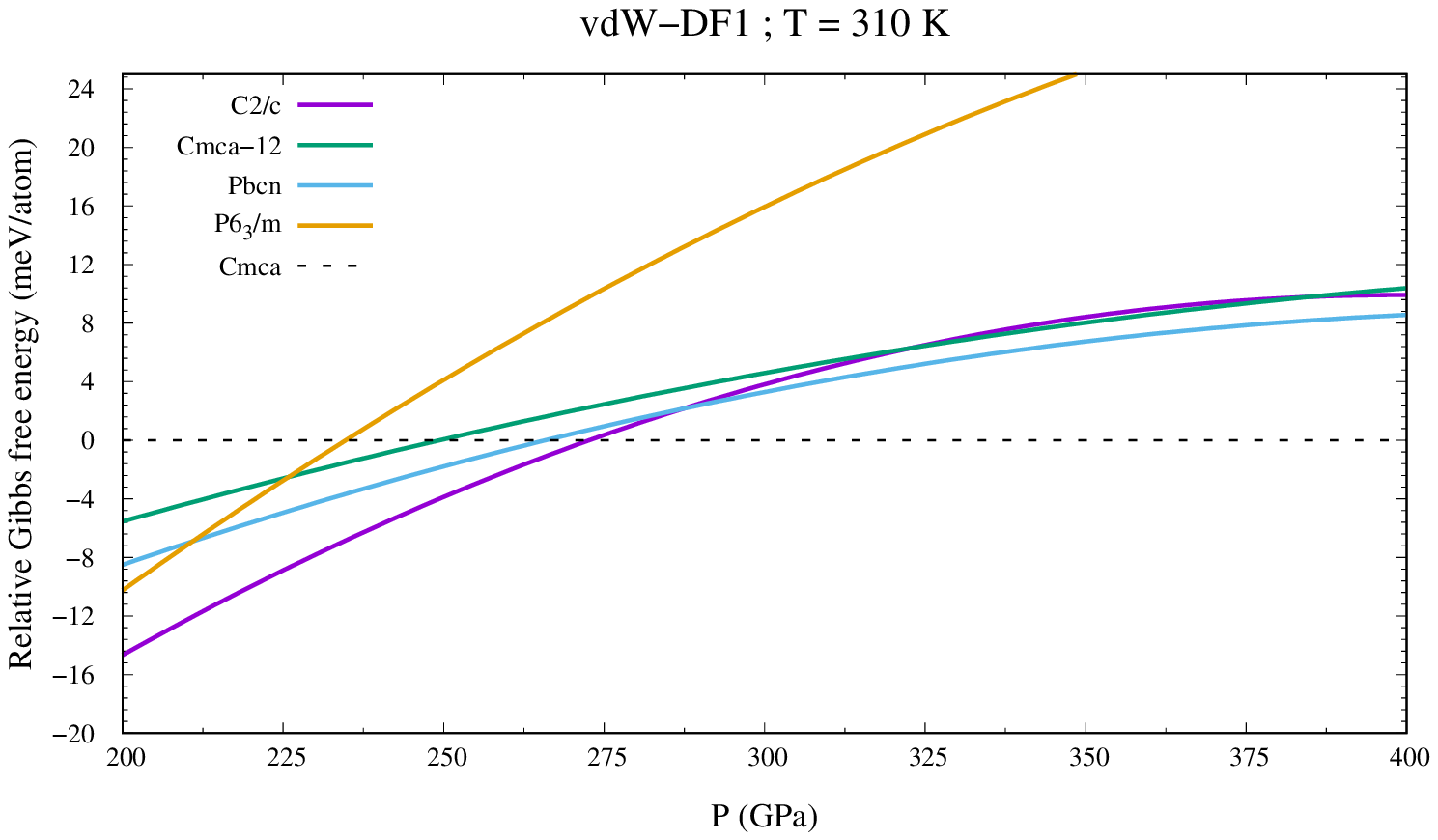}&
\includegraphics[width=0.5\textwidth]{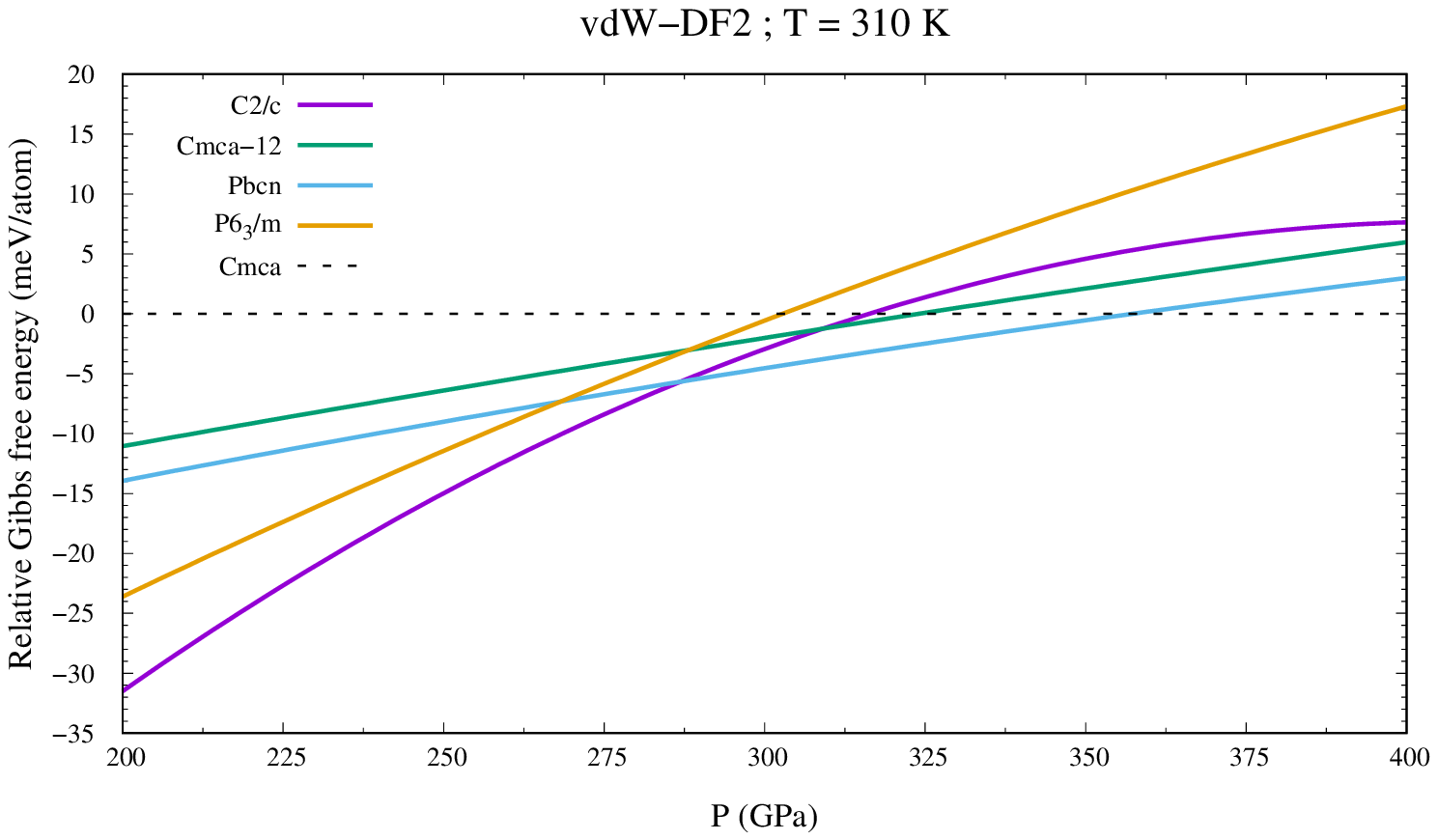}\\
\end{tabular}
\caption{\label{gibbs} (Color online) Relative Gibbs free energy per atom as a function of
pressure at T = 10, and 310 K calculated using vdW-DF1 and vdw-DF2.
The Gibbs free energy of molecular crystal structures are presented
relative to the metallic $Cmca$ structure.} 
\end{figure}

We calculated the quasiharmonic Gibbs free energy for high pressure
solid molecular hydrogen phases utilising the vdW functionals. Figure
\ref{gibbs} illustrates the Gibbs free energy as a function of
pressure at T = 10, and 310 K.  Lattice vibrations and zero point (ZP)
contributions play a crucial role in determining phase boundaries.
Our low-temperature vdW-DF1 phase diagram is rather similar to
previous work with PBE.  It predicts that the $C2/c$ is the most
stable phase up to 289 GPa where it transforms to metallic $Cmca$
phase. Increasing the temperature to 310 K reduces the $C2/c$ to
$Cmca$ phase transition pressure to 275 GPa. Comparing to previous DFT
phase diagram results, this is the lowest DFT molecular insulator to
molecular metallic phase transition pressure. vdW-DF1 phase diagram
predicts that metallization of high-pressure solid hydrogen occurs
through molecular-molecular phase transition at pressure below 300
GPa, whereas experiments \cite{Howie,Howie2,zha,loubey,silvera}
suggest that the metallization of solid
hydrogen takes place at pressures larger than 350 GPa. 
Including the lattice dynamic contribution to the vdW-DF1 phase diagram 
increases the discrepancy between vdW-DF1 outcomes and experimental
observations as well as our recently reported QMC results\cite{PRB17}.

Our vdW-DF2 Gibbs free energy calculations (figure \ref{gibbs})
indicate two low temperature phase transitions of $C2/c$ to $Pbcn$ at
291 GPa and $Pbcn$ to $Cmca$ at 402 GPa.  Whatever functional is
used, the effects of nuclear quantum and thermal vibrations play a
crucial role in the stabilization of phase III and IV.  Recent DFT
calculations report a new hexagonal structure with $P6_122$ symmetry
for phase III of solid molecular hydrogen, which is more stable than
$C2/c$ at pressures below 200 GPa \cite{monserrat16}.  We suggest that
two molecular insulator structures with molecular positions close to
hcp could be stabilized above 200 GPa in the region ascribe to
phase III: monoclinic $C2/c$ up to
291 GPa, and $Pbcn$ up to 402 GPa pressure. By increasing the
temperature to T = 310 K the insulator $Pbcn$ to metallic $Cmca$ phase
transition occurs at 366 GPa. Our previous QMC results\cite{samprl}
predict that molecular to atomic phase transition takes place at about
374 GPa. Our recent quasi-particle and excitonic band gap
study\cite{PRB17} also suggests that band-gap closure of best
candidates for solid molecular structures occurs within pressure range
of 350-400 GPa. This work predicts that insulator to metallic phase
transition happens at 366 GPa. Based on our extensive study of the
metallisation of high pressure solid hydrogen, we conclude that all
three scenarios of metallisation, which are molecular-atomic
structural transformation, band-gap closure and insulator molecular to
metallic molecular phase transition, indicate that solid hydrogen
become a metal at pressure range of 350-400 GPa. This prediction
agrees well with experiments\cite{Howie,Howie2,loubey}.

\begin{figure}
\centering
\includegraphics[width=0.75\textwidth]{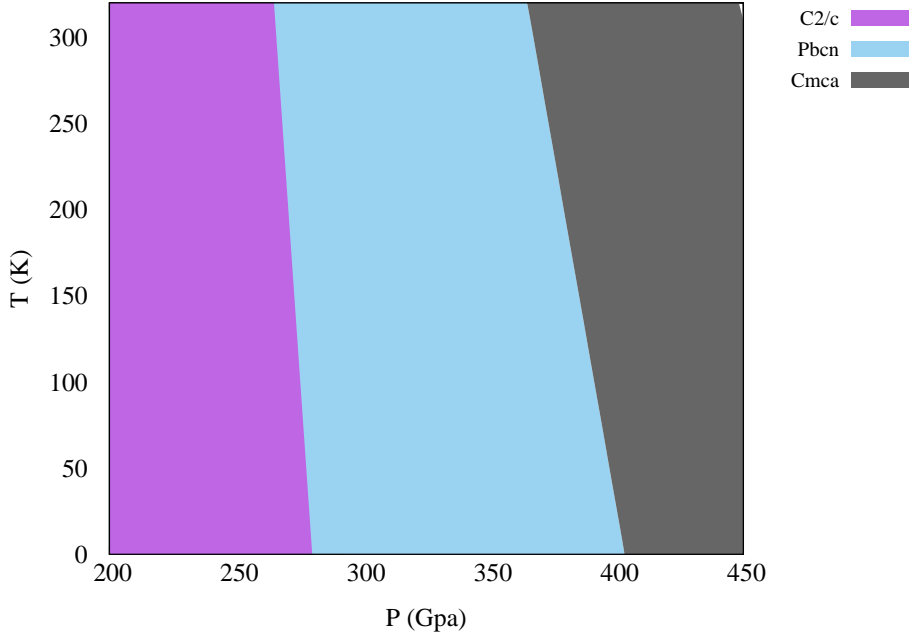}
\caption{\label{TP} (Color online) Predicted
    temperature-pressure phase diagram for solid molecular hydrogen
    obtained by vdW-DF2. The zero point fluctuations are the
    influential term in the Gibbs free energy calculations which
    stabilises the $Pbcn$ and metallic $Cmca$ phases. The main
    contribution to zero-point comes from the vibrons, while the
    thermal-phonon effects depend more on the lattice modes}
\end{figure}

 Figure \ref{TP} illustrates temperature-pressure phase
  diagram for solid molecular hydrogen which is predicted by
  vdW-DF2. The metallisation transition to $Cmca$ is strongly affected by  
  quantum zero point fluctuations and 
  occurs above 400GPa,consistent with experiment.  Zero point fluctuations also destabilize
  $C2/c$ with respect to $Pbcn$ around 300GPa with either functional.  Increasing
  the temperature reduces the metal-insulator phase transition pressure.
  The stability of phase III, at room temperature has been observed 
  experimentally\cite{Howie}. The Raman and visible transmission spectroscopy 
  measurements at 300 K and up to 315 GPa indicate the phase transformation 
  to phase III around 200 GPa. 

 Finally we argue that the vdW-DF2 functional provides
better results at high density limit than vdW-DF1 for the same reason
that BLYP performs better than PBE. The GGA functional, which is used
in both vdW-DF1 and vdW-DF2, can be given by
\begin{equation}
 E_x^{GGA}[n({\bf r}), \nabla n({\bf r})] = -(3/4)(3/\pi)^{1/3} \int d^3 r n^{4/3} A_x^{GGA}(s)
\end{equation}
where $s = (\nabla n)/(2k_Fn)$, and $k_F = (3\pi^2n)^{1/3}$ is the local Fermi wave vector.
The PBE and revPBE enhancement factor formula , which is used in vdW-DF1, is
\begin{equation}
 A_x(s) = 1 + \frac{\mu s^2}{1+\mu s^2/\kappa}
\end{equation}
where both PBE and revPBE use $\mu = 0.2195$ which correctly describes
the low $s$ limit but  and PBE and revPBE become
insensitive to $s$ in the high-$s$ limit.  For $H_2$ dimers, significant
values of $s$ as large as 25 is obtained\cite{Emmon},
which can yield spurious exchange attractions in PBE and vdW-DF1.

The PW86, which is almost linear in $s$, gives an enhancement factor
proportional to $s^{2/5}$ at large $s$, and provides net repulsive
interaction for exchange energy\cite{Kannemann}.  Detailed analysis of
GGA functionals\cite{Emmon,Kannemann} indicate that PW86 (as used in
vdW-DF2) is the best for systems dominated by large $s$ such as high
pressure hydrogen.
\section{Conclusion}\label{CON}
We have employed non-local vdW functionals to revisit
the phase diagram of high-pressure solid hydrogen within 
pressure range of $200 < P < 450$ GPa. 
We studied the best candidates for
phase III previously discovered by structure searching using the PBE
approach. In phase III the H$_2$ bond weakens with
pressure as electrons delocalize, and there is competition between
insulating and metallic, molecular and atomic structure.
Consequently, the cancellation of exchange-correlation errors which
typically allows DFT to give accurate energy differences is most
sorely tested here. In previous work it was shown that, compared with
experiment and QMC calculation, PBE obtains over-long BLs, too-low
vibron frequencies, and too-low transition pressure to the molecular metallic
phase ($Cmca$).  By considering a range of XC functionals, we have
shown that these failings are all related.

To examine the importance of the long-range vdW interactions in 
solid molecular structures of high-pressure hydrogen,
 we made use of two widely applied vdW functionals of vdW-DF1 and 
vdW-DF2 to calculate static-enthalpy
and finite-temperature dynamic Gibbs free energy as functions of
pressure.  The vdW-DF1 gives erroneous results similar to PBE, while vdW-DF2
gets bond lengths, frequencies and transition pressures close to QMC
and experimental results.  

The distinguishing feature of the more successful functionals is the
treatment of semi-local exchange rather than in inclusion of vdW.  
In particular those functionals
which correctly fit the limit of high charge density gradient give
better-defined molecules.  Interestingly, this effect is most
pronounced in the molecular metallic phase ($Cmca$) which is overly
stable in PBE due to the low vibron frequency and consequent low
zero-point energy.

The sensitivity of the metallization pressure to choice of XC
functional is likely to be a feature of all hydrogen phases.  Our
results suggest that previous DFT calculations of metallization
pressure, including the metallization of the liquid, will have an
uncertainty of order $\pm$100 GPa, with the widely used PBE functional
giving especially low values. 

\section{Acknowledgments}
This work was supported by the European Research Council (ERC) Grant
"Hecate" reference No. 695527.  Computing facilities were provided
through DECI-13 PRACE project "QMCBENZ15" and the Dutch national
supercomputer Cartesius.  S. Azadi acknowledges useful discussions
with \u{E}. D. Murray. GJA acknowledges support from
  EPSRC (UKCP grant K01465X) and a Royal Society Wolfson fellowship.

\end{document}